\documentclass[a4paper,10pt,twoside]{cpc-hepnp}
\usepackage{CJK,upgreek,fancyhdr}
\usepackage{multicol}
\usepackage{graphicx}
\usepackage{booktabs}
\usepackage{amssymb,bm,mathrsfs,bbm,amscd}
\usepackage[tbtags]{amsmath}
\usepackage{lastpage}
\usepackage{hyperref}

\begin{document}
\begin{CJK*}{GB}{gbsn}
\title{Global Constraints from RHIC and LHC on Transport Properties of QCD Fluids in CUJET/CIBJET Framework} 
\author{
		Shuzhe Shi$^{1;1}$\email{shishuz@indiana.edu} 
\quad	Jinfeng Liao$^{1;2}$\email{liaoji@indiana.edu} 
\quad	Miklos Gyulassy$^{2,3,4;3}$\email{mg150@columbia.edu}
}
\maketitle
\address{$^1$ Physics Department and Center for Exploration of Energy and Matter,
Indiana University, 2401 N Milo B. Sampson Lane, Bloomington, IN 47408, USA.}
\address{$^2$ Nuclear Science Division, Lawrence Berkeley National Laboratory, Berkeley, CA 94720, USA.}
\address{$^3$ Pupin Lab MS-5202, Department of Physics, Columbia University, New York, NY 10027, USA.}
\address{$^4$ Institute of Particle Physics and Key Laboratory of Quark \& Lepton Physics (MOE), Central China Normal University, Wuhan, 430079, China.}

\date{\today}

\begin{abstract}
We report results of a comprehensive global $\chi^2$ analysis of nuclear collision data from RHIC (0.2 ATeV), LHC1 (2.76 ATeV), and recent LHC2 (5.02 ATeV) energies using the updated CUJET framework. The framework consistently combines viscous hydrodynamic fields predicted by VISHNU2+1 (validated with soft $p_T<2$~GeV bulk observables) and the DGLV theory of jet elastic and inelastic energy loss generalized to QGP fluids with an sQGMP color structure, including effective semi-QGP color electric quark and gluon as well as emergent  color magnetic monopole degrees of freedom constrained by lattice QCD data. We vary the two control parameters of the model (the maximum value of the running QCD coupling, $\alpha_c$, and the ratio $c_m$ of color magnetic to electric screening scales) and  calculate the global $\chi^2(\alpha_c,c_m)$ compared with available jet fragment observables ($R_{AA}, v_2$). A global $\chi^2<2$ minimum is found with $\alpha_c \approx 0.9\pm 0.1$ and $c_m\approx 0.25\pm 0.03$. Using CIBJET, the event-by-event (ebe) generalization of the CUJET framework, we show that ebe fluctuations in the initial conditions do not significantly alter our conclusions (except for $v_3$). An important theoretical advantage of the CUJET and CIBJET frameworks is not only its global $\chi^2$ consistency with jet fragment observables at RHIC and LHC and with non-perturbative lattice QCD data, but also its internal consistency of the constrained jet transport coefficient, $\hat{q}(E,T)/T^3$, with the near-perfect fluid  viscosity to entropy ratio ($\eta/s \sim T^3/\hat{q}\sim 0.1-0.2$) property of QCD fluids near $T_c$ needed to account for the low $p_T<2$ GeV flow observables.  Predictions for future tests at LHC with 5.44 ATeV Xe + Xe and 5.02 ATeV Pb + Pb are also presented.
\end{abstract}

\begin{keyword}
jet quenching, quark-gluon plasma, relativistic heavy-ion collisions, heavy flavor physics
\end{keyword}
\begin{pacs}
12.38.Aw, 12.38.Mh
\end{pacs}
\newpage

\section{Introduction}
High energy quark and gluon jets, initially generated in rare perturbative QCD processes, lose energy and diffuse transversely along their paths due to interactions with microscopic constituents in the hot quark-gluon plasma created by heavy ion collisions at the Relativistic Heavy Ion Collider (RHIC) and the Large Hadron Collider (LHC).   
Such hard ($p_T >10$~GeV) processes provide an independent probe of the evolution history of the soft QCD matter ($p_T<2$~GeV) produced in such collisions. 
Recent high-precision data from LHC Pb + Pb collisions  on jet quenching and azimuthal asymmetry observables over wide kinematics and centrality ranges provide an opportunity to quantitatively constrain and differentiate competing models of jet-medium interactions, as well as varied assumptions  of the chromo-electric and magnetic field structure of the bulk QCD ``perfect fluids'' produced in ultra-relativistic nuclear collisions.

Given (i) a detailed microscopy theory of jet medium interactions ({\it e.g.} DGLV~\cite{Gyulassy:1993hr,Gyulassy:1999zd,Gyulassy:2000er,Gyulassy:2002yv,Vitev:2002pf,Djordjevic:2003zk,Gyulassy:2003mc,Wicks:2005gt,Buzzatti:2011vt}, HT~\cite{Guo:2000nz,Wang:2001ifa,Majumder:2007ae,Cao:2017hhk}, AMY~\cite{Arnold:2002ja,Arnold:2002zm,Arnold:2003zc}, or AdS~\cite{Liu:2006ug,Liu:2006he}), (ii) a detailed model of bulk initial conditions ({\it e.g.} Glauber~\cite{Miller:2007ri}, TRENTO~\cite{Moreland:2014oya}, or CGC~\cite{Iancu:2003xm}),  and (iii)
a long wavelength collective transport theory of the bulk QCD matter, such as relativistic viscous hydrodynamics ({\it e.g.} VISHNU~\cite{Shen:2010uy},  vUSPHydro~\cite{Noronha-Hostler:2013gga,Noronha-Hostler:2014dqa,Noronha-Hostler:2015coa}, or MUSIC~\cite{Schenke:2010nt,Schenke:2010rr}), the observed attenuation pattern of hard jet observables and their correlations with soft bulk collective flow observables can help differentiate competing dynamical models of high energy A + A collisions.
In Refs.~\cite{Xu:2014tda,Xu:2015bbz} we developed the CUJET3.0 framework that combines the DGLV theory of jet energy loss coupled with nearly ``perfect QCD fluids'' described by the viscous hydrodynamics theory (and simulated via VISHNU~\cite{Shen:2014vra}) to constrain the color degrees of freedom.

\begin{figure}[!hbt]\centering
\includegraphics[width=0.53\textwidth]{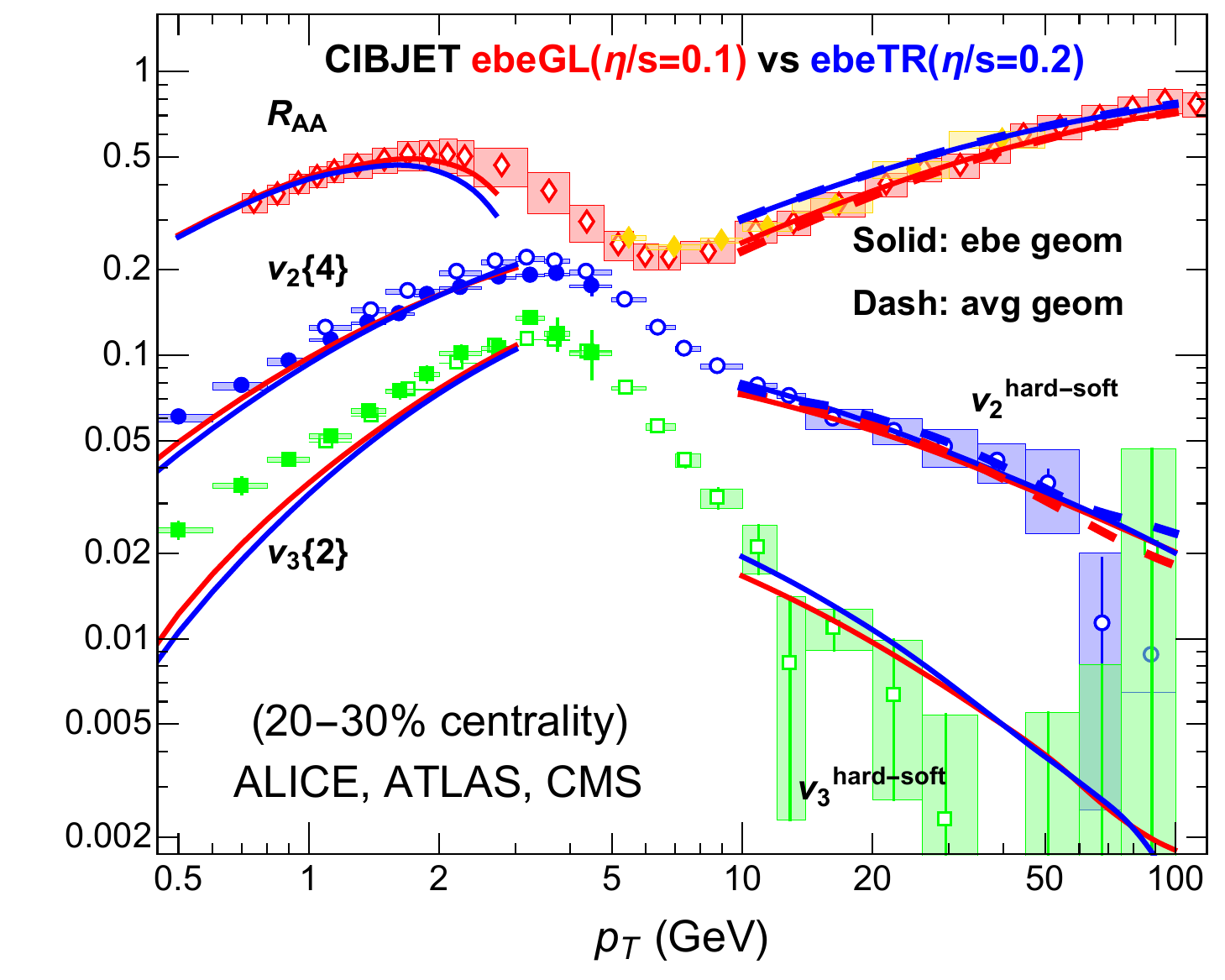}
\caption{(color online) Nuclear modification factor $R_{AA}$ as well as the second and third harmonic coefficients $v_2$ and $v_3$ of final hadron azimuthal distribution as functions of $p_T$ for 20--30\% Pb + Pb collisions at 5.02~ATeV. Solid curves are obtained from event-by-event calculations, while the dashed curves depict averaged smooth geometry. CIBJET results in both soft and hard regions, with either Monte-Carlo Glauber (red)  or Trento (blue) initial conditions, are in excellent agreement with experimental data from  ALICE, ATLAS, and CMS~\cite{Acharya:2018qsh,ATLAS:2017rmz,Khachatryan:2016odn,Sirunyan:2017pan,Adam:2016izf}. Similar CIBJET results for 30--40\% centrality, in excellent agreement with experimental data, were shown in ~\cite{Shi:2018lsf}.
}
\label{fig.cibjet}
\end{figure}

The simplest class of hard observables in a specific centrality class, ${\cal C}$,  is the $p_T$ and relative azimuthal angle
dependence of the nuclear modification factor $R_{AA}^f$ for final state hadrons (with flavor species denoted by $f$), which is decomposed into Fourier harmonics as:
\begin{eqnarray}\begin{split}
  R_{AA}^f\left (p_T,\phi; {\cal C}; \sqrt{s}\right ) =&\; \frac{\frac{dN_{AA}^f({\cal C})}{p_Tdp_Td\phi}}{T_{AA}({\cal C}) \frac{d\sigma_{pp}^f}{p_Tdp_Td\phi}} \\
  =&\;  R_{AA}^f\left (p_T; {\cal C}; \sqrt{s}\right ) \left[ 1 + 2 \sum_n \langle
    v_n^f\left (p_T; {\cal C}; \sqrt{s}; \right ) \cos(n (\phi- \Psi_n))\rangle_{\cal C}\right]
\end{split}\end{eqnarray}
where $T_{AA}({\cal C})$ is the average number of binary nucleon-nucleon scattering per unit area in centrality class ${\cal C}$. 
Typically, ${\cal C}$ is expressed as a percentage interval of the inelastic cross section, {\it e.g.} 10 -- 20\% of the charged multiplicity per unit rapidity distribution.
The  $p_T$ and $\phi$ depict the transverse momentum and the azimuthal angle of observed leading hadrons, respectively, relative to the bulk collective flow azimuthal harmonics.
The experimental measurements of hard particle harmonics $v_n^f$ are performed with respect to event-wise soft harmonics, and event-by-event fluctuations of the bulk initial condition may play an important role~\cite{Noronha-Hostler:2016eow}. Within the CUJET3 framework, the influence of event-by-event fluctuations has been investigated with a generalized CIBJET (= ebeIC + VISHNU + DGLV) framework, with the results reported in  Ref.~\cite{Shi:2018lsf}.  The  CIBJET results of $R_{AA}$, $v_2$, and $v_3$ observables across a very wide range of $p_T$ for  30--40\% centrality Pb + Pb collisions at 5.02~ATeV were shown in ~\cite{Shi:2018lsf}, and they are in excellent agreement with experimental data. In Fig.~\ref{fig.cibjet}, we further present the CIBJET results of  $R_{AA}$, $v_2$ and $v_3$ for a different centrality of 20--30\%,  which likewise show excellent agreement with experimental data and demonstrate the correct centrality dependence of the CIBJET results. One conclusion found with CIBJET is that the $p_T$ and centrality dependence of the elliptic $v_2^f(p_T,{\cal C})$ azimuthal harmonics shows quantitative consistency at a $\sim 10\%$ level between calculations with averaged smooth bulk geometry and those with fluctuating initial conditions. 
This conclusion is true for the varied centrality class and is in agreement with  a similar consistency-check from the ebeIC + LBT + HT hard + soft framework in Ref.~\cite{Cao:2017umt,Cao:2017hhk}, while different from the ebeIC + vUSPhydro + BBMG framework in Ref.~\cite{Noronha-Hostler:2016eow}, which found a much larger sensitivity (factor $\sim 2$) of the hard elliptic harmonic to event-by-event fluctuations. The finding from CIBJET justifies the use of averaged smooth geometry in the CUJET3 framework, as we shall adopt in the present paper.  

The prime motivation of this work is to conduct a comprehensive new global $\chi^2$ analysis of nuclear collision data from RHIC (0.2ATeV), LHC1 (2.76ATeV), and recent LHC2 (5.02ATeV) energies for high $p_T$ light and heavy flavor hadrons. This analysis is performed with the updated CUJET3.1 framework to  evaluate jet energy loss distributions in various models of the color structure of QCD fluids produced in heavy ion collisions. The CUJET3.1 is based on our previous CUJET3.0 framework~~\cite{Xu:2014tda,Xu:2015bbz}  and successfully addressed a few issues in CUJET3.0. A brief introduction to CUJET3.0 and a detailed discussion regarding the improvements in CUJET3.1 are included in the two appendices. We will show that CUJET3.1 provides a non-perturbative solution to the long standing hard ($R_{AA}$ and $v_2$) versus soft ``perfect fluidity'' puzzle. We further examine the crucial issue of consistency between soft and hard transport properties of the QCD fluid in this framework. Predictions for future tests at LHC with 5.44 ATeV Xe + Xe and 5.02 ATeV Pb + Pb will  also be presented.

The organization of this paper is as follows.  
We perform the model parameter optimization in Sec.~\ref{sec.calibration}, based on the quantitative $\chi^2$ analysis with a comprehensive set of experimental data for light hadrons. 
In Sec.~\ref{sec.comparison}, we show the successful CUJET3.1 description of available experimental data for light hadrons as well as the successful independent test with heavy flavor hadrons. 
The temperature dependence of the jet transport coefficient and the corresponding shear viscosity for the quark-gluon plasma, extracted from CUJET3.1,  are presented  in Sec.~\ref{sec.coef}. 
CUJET3.1 predictions for on going experimental analysis are shown in Sec.~\ref{sec.prediction}. 
Finally, we summarize the paper in Sec.~\ref{sec.summary}.
A brief introduction of the CUJET3 framework, as well as the improvements made in CUJET3.1, are included in the two appendices.  

\section{Global $\chi^2$ Analysis with CUJET3}\label{sec.calibration}

As discussed in Appendix~\ref{sec.cujet3}, the CUJET3 framework is a quantification model solving jet energy loss in a hydrodynamics background, implementing DGLV jet energy loss from both inelastic and elastic scattering, and taking into account interactions with both chromo-electric and magnetic charges of the medium.
There are two key parameters in the model. One is $\alpha_c$ (see also Eq.\ref{TcEnhancement} in App.\ref{sec.cujet3}):
\begin{equation*}
\alpha_s(Q^2) = \frac{\alpha_c}{1+\frac{9\alpha_c}{4\pi}\log(Q^2/T_c^2)}\,,
\end{equation*}
which is the value of QCD running coupling at the non-perturbative scale $Q^2 = T_c^2$. It sensitively controls and positively influences the overall opaqueness of the hot medium. 
The other is $c_m$, defined via $\mu_M = c_m\,g(T)\,\mu$, (see also Eq.\ref{TcEnhancement}, \ref{DebyeMass} and \ref{f_EM} in App.\ref{sec.cujet3}), which is the coefficient for magnetic screening mass in the medium and influences the contribution of the magnetic component to the jet energy loss. The increase of $c_m$ leads to the enhancement of monopole mass, hence overall opaqueness. 
Magnetic mass scales with magnetic scale $g^2T$, but its coefficient receives non-perturbative contributions and can-not be perturbatively calculated even at high temperature.
Constrained by the lattice QCD calculation~\cite{Nakamura:2003pu}, the reasonable value of $c_m$ varies in the range of $0.2\lesssim c_m \lesssim0.5$.

To systematically constrain these two key parameters, first we perform a quantitative $\chi^2$ analysis and utilize central and semi-central high transverse momentum light hadron's $R_{AA}$ and $v_2$ for all available data.
We compare the relative variance between theoretical expectation and experimental data, which is defined as the ratio of squared difference between experimental data points and corresponding CUJET3 expectation, to the quadratic sum of experimental statistic and systematic uncertainties for that data point:
\begin{equation}
\chi^2/\mathrm{d.o.f.} = \sum_i \frac{(y_{\mathrm{exp},i} - y_{\mathrm{theo},i})^2}{\sum_{s}(\sigma_{s,i})^2} \Bigg/ \sum_i 1,
\end{equation}
where $\sum_i$ runs over all experimental data point in the momentum range $8 \leq p_T \leq 50$~GeV/$c$, and $\sum_s$ denotes summing over all sources of uncertainties, e.g. systematic and statistic uncertainties.
We compute $\chi^2/\mathrm{d.o.f.}$ for each of the following 12 data sets:\\
$\bullet$\quad 200~GeV Au-Au Collisions, 0--10\% Centrality Bin, $R_{AA}(\pi^0)$: PHENIX~\cite{Adare:2008qa,Adare:2012wg};\\
$\bullet$\quad 200~GeV Au-Au Collisions, 0--10\% Centrality Bin, $v_{2}(\pi^0)$: PHENIX~\cite{Adare:2012wg};\\
$\bullet$\quad 200~GeV Au-Au Collisions, 20--30\% Centrality Bin, $R_{AA}(\pi^0)$: PHENIX~\cite{Adare:2008qa,Adare:2012wg};\\
$\bullet$\quad 200~GeV Au-Au Collisions, 20--30\% Centrality Bin, $v_{2}(\pi^0)$: PHENIX~\cite{Adare:2012wg};\\
$\bullet$\quad 2.76~TeV Pb-Pb Collisions, 0--10\% Centrality Bin, $R_{AA}(h^{\pm})$: ALICE~\cite{Abelev:2012hxa};\\
$\bullet$\quad 2.76~TeV Pb-Pb Collisions, 0--10\% Centrality Bin, $v_{2}(h^{\pm})$: ATLAS~\cite{ATLAS:2011ah}, CMS~\cite{Chatrchyan:2012xq};\\
$\bullet$\quad 2.76~TeV Pb-Pb Collisions, 20--30\% Centrality Bin, $R_{AA}(h^{\pm})$: ALICE~\cite{Abelev:2012hxa};\\
$\bullet$\quad 2.76~TeV Pb-Pb Collisions, 20--30\% Centrality Bin, $v_{2}(h^{\pm})$: ALICE~\cite{Abelev:2012di}, ATLAS~\cite{ATLAS:2011ah}, CMS~\cite{Chatrchyan:2012xq};\\
$\bullet$\quad 5.02~TeV Pb-Pb Collisions, 0--5\% Centrality Bin, $R_{AA}(h^{\pm})$: ATLAS-preliminary~\cite{ATLAS:2017rmz}, CMS~\cite{Khachatryan:2016odn};\\
$\bullet$\quad 5.02~TeV Pb-Pb Collisions, 0--5\% Centrality Bin, $v_{2}(h^{\pm})$:  CMS~\cite{Sirunyan:2017pan};\\
$\bullet$\quad 5.02~TeV Pb-Pb Collisions, 10--30\% Centrality Bin, $R_{AA}(h^{\pm})$: CMS~\cite{Khachatryan:2016odn};\\
$\bullet$\quad 5.02~TeV Pb-Pb Collisions, 20--30\% Centrality Bin, $v_{2}(h^{\pm})$: CMS~\cite{Sirunyan:2017pan};\\
and finally obtain the overall $\chi^2/\mathrm{d.o.f.}$ as the average over these data sets.

\begin{figure}[!hbt]\centering
\includegraphics[width=0.95\textwidth]{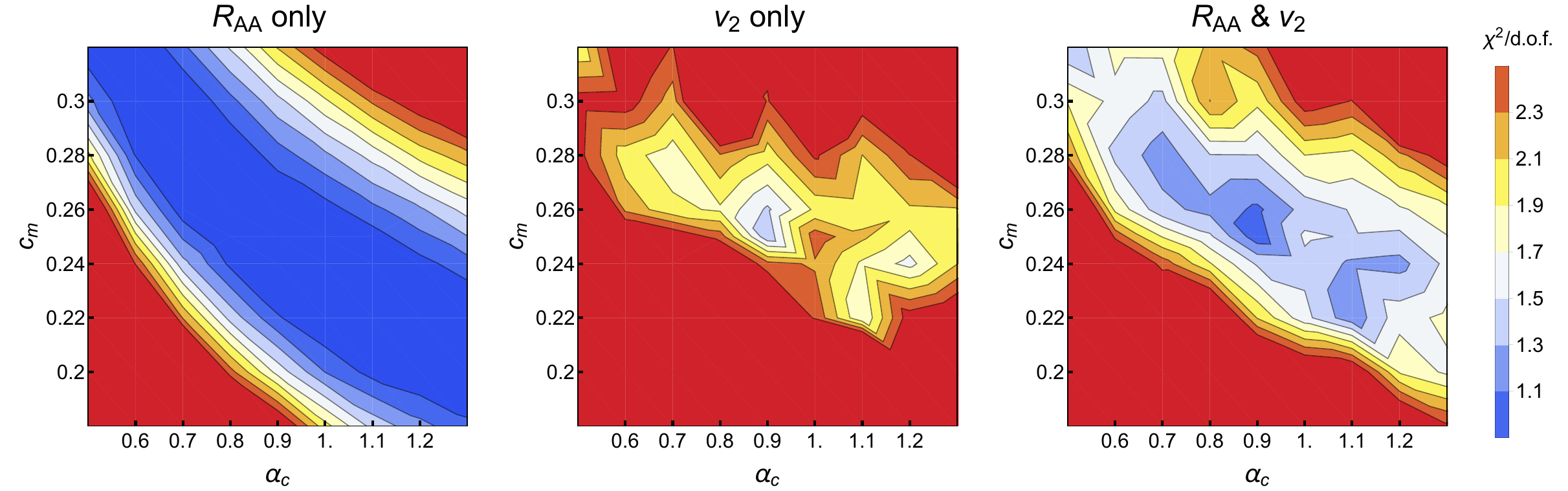}
\caption{(color online) $\chi^2/\mathrm{d.o.f.}$ comparing $\chi_T^L$-scheme CUJET3 results with RHIC and LHC data. 
Left: $\chi^2/\mathrm{d.o.f.}$ for $R_{AA}$ only.
Middle: $\chi^2/\mathrm{d.o.f.}$ for $v_2$ only.
Right: $\chi^2/\mathrm{d.o.f.}$ including both $R_{AA}$ and $v_2$}
\label{fig.ChiSq_LE_full}
\end{figure}
First of all, we perform the analysis in the ``slow'' quark-libration scheme ($\chi_T^L$-scheme) for a wide range of parameter space: $0.5 \leq \alpha_c \leq 1.3$, $0.18 \leq c_m \leq 0.32$.
As shown in Fig.~\ref{fig.ChiSq_LE_full}, $\chi^2/\mathrm{d.o.f.}$ with only $R_{AA}$ data (left panel) or only $v_2$ data (middle panel) yields different tension and favors different regions of parameter space.
Taking all data together (right panel), we identify a data-selected optimal parameter set as ($\alpha_c = 0.9$, $c_m = 0.25$), with $\chi^2/\mathrm{d.o.f.}$ close to unity,
while the ``uncertainty region'' spanned by ($\alpha_c = 0.8$, $c_m = 0.22$) and ($\alpha_c = 1.0$, $c_m = 0.28$) with a $\chi^2/\mathrm{d.o.f.}$ about two times the minimal value.
Both the optimal parameter set and the ``uncertainty region'' remain essentially unchanged if $\chi^2/\mathrm{d.o.f.}$ is computed giving the same weight for each data point instead of each data set.

\begin{figure}[!hbt]\centering
\includegraphics[width=0.99\textwidth]{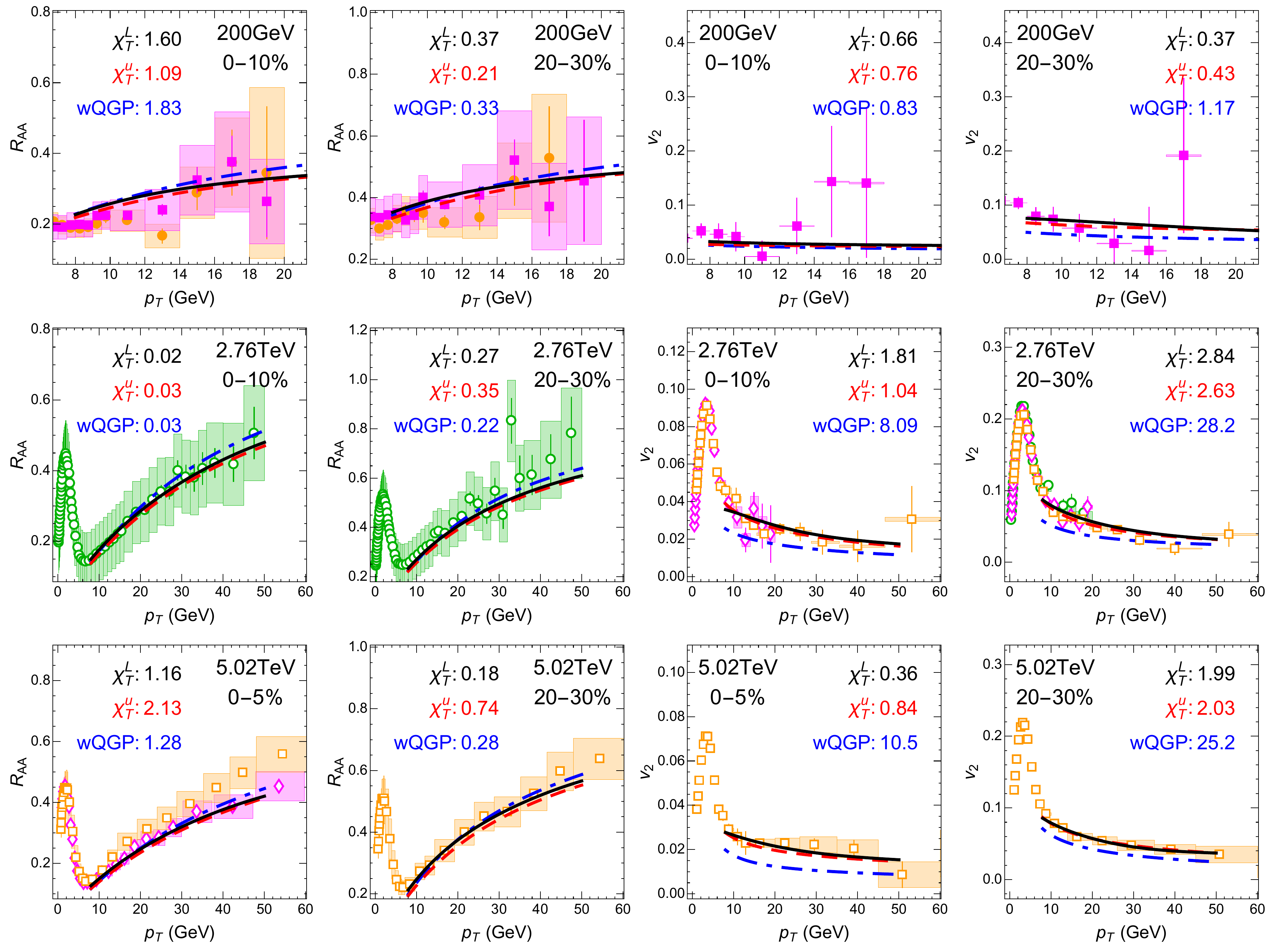}
\caption{(color online) CUJET theoretical expectation of light hadron $R_{AA}$ and $v_2$ using three different schemes: sQGMP $\chi_T^L$-scheme (black solid), sQGMP $\chi_T^u$-scheme (red dashed), wQGP/CUJET2 scheme (blue dashed dotted). Corresponding $\chi^2/\mathrm{d.o.f.}$ are shown, with respect to following experimental data: PHENIX 2008 (orange solid circle)~\cite{Adare:2008qa}, PHENIX 2012 (magenta solid square)~\cite{Adare:2012wg}; ALICE (magenta open diamond)~\cite{Abelev:2012hxa,Abelev:2012di}, ATLAS (green open circle)~\cite{ATLAS:2011ah,ATLAS:2017rmz}, CMS (orange open square)~\cite{Chatrchyan:2012xq,Khachatryan:2016odn,Sirunyan:2017pan}. }
\label{fig.full}
\end{figure}
In order to test the need of the of chromo-magnetic-monopole (cmm) degrees of freedom and to explore the potential influence of theoretical uncertainties of different quark liberation schemes, we perform the same $\chi^2$ analysis with two other schemes: (a) the ``fast'' quark-libration scheme ($\chi_T^u$-scheme); (b) the weakly coupling QGP~(wQGP) scheme, equivalent to CUJET2.0 mode, and assuming no cmm, i.e. taking $f_E=1$, $f_M=0$, and chromo-electric-components fraction $\chi_T=1$, while the running coupling takes the Zakharov formula as in Eq.~(\ref{AlphaRunMax}).

By using these three schemes with their corresponding most optimal parameter set:\\
$\bullet$\quad (i) sQGMP $\chi_T^L$-scheme: $\alpha_c = 0.9$, $c_m = 0.25$,\\
$\bullet$\quad (ii) sQGMP $\chi_T^u$-scheme: $\alpha_c = 0.9$, $c_m = 0.34$,\\
$\bullet$\quad (iii) wQGP/CUJET2 scheme: $\alpha_{\mathrm{max}} = 0.4$, (optimized by $R_{AA}$)\\
we show in Fig.~\ref{fig.full} their comparison with above the experimental data sets, including the quantitative value of $\chi^2/\mathrm{d.o.f.}$ for each data set.
While both sQGMP schemes ($\chi_T^L$ and $\chi_T^u$) give similar jet quenching variables, the QGP scheme gives similar $R_{AA}$ but less azimuthal anisotropy. In particular, one can see clearly from the quantitative value of their $\chi^2/\mathrm{d.o.f.}$ that the theoretical expectations of both sQGMP schemes are in good consistency with the experimental data, and that of the QGP scheme, without cmm degree of freedom, differs significantly from the highly precise LHC $v_2$ measurements.
The $\chi^2$ analysis strongly supports the need of cmm degrees of freedom, but remains robust on the specific quark liberation scheme.

While we maintain the unification of the CUJET3 model by using the same (globally optimized) parameter set, it's worth mentioning that quantitative $\chi^2$ analysis for a different data set, e.g. a different observable or different beam energy, flavors a different parameter regime, as shown in Tab.~\ref{tab.ChiSq}.
When comparing to the $R_{AA}$ results, the azimuthal anisotropy measurement with more shrink uncertainties yields higher $\chi^2/\mathrm{d.o.f}$ and hence has stronger constrain power.
On the other hand, in CUJET3 models, the RHIC results flavor stronger coupling (larger $\alpha_c$ or $c_m$) than the LHC results.
Meanwhile, the latter are more precise and provide better distinction of different models.
Particularly in the case of 5.02~TeV data, the sQGMP schemes are explicitly more phenomenologically flavored than the wQGP scheme.
\begin{table}[!hbt]\centering
\begin{tabular}{c|cp{4mm}cp{4mm}c|cp{4mm}c|cp{4mm}c}
\hline\hline
		& \multicolumn{5}{c}{sQGMP $\chi_T^L$} & \multicolumn{3}{|c|}{sQGMP $\chi_T^u$} & \multicolumn{3}{c}{wQGP} \\
\cline{2-12}
		& $\quad\alpha_c\quad$ && $c_m$ && $\chi^2/\mathrm{d.o.f}$ & $\quad c_m\quad$ && $\chi^2/\mathrm{d.o.f}$ & $\quad\alpha_{\mathrm{max}\quad}$ && $\chi^2/\mathrm{d.o.f}$ \\
\hline
$R_{AA}$	& 0.9 && 0.24 && 0.57	&	0.31 && 0.60	&	0.4 && 0.67 \\
$v_2$	& 0.9 && 0.25 && 1.34	&	0.34 && 1.28	&	1.0 && 2.34 \\
\hline
200~GeV	& 1.2 && 0.28 && 0.40	&	0.40 && 0.42	&	0.6 && 0.61 \\
2.76~TeV	& 0.9 && 0.24 && 1.15	&	0.34 && 1.01	&	1.0 && 2.07 \\
5.02~TeV	& 0.7 && 0.28 && 0.76	&	0.34 && 1.43	&	1.0 && 8.61 \\
\hline
All		& 0.9 && 0.25 && 0.97	&	0.34 && 1.02	&	0.7 && 3.47	\\
\hline
\hline
\end{tabular}
\caption{Optimal parameter and corresponding $\chi^2/\mathrm{d.o.f.}$ for different data sets in different schemes. 
Note that the sQGMP $\chi_T^u$ scheme is optimized by taking $\alpha_c\equiv0.9$.
\label{tab.ChiSq}}
\end{table}

\begin{figure}[!hbt]\centering
\includegraphics[width=0.3\textwidth]{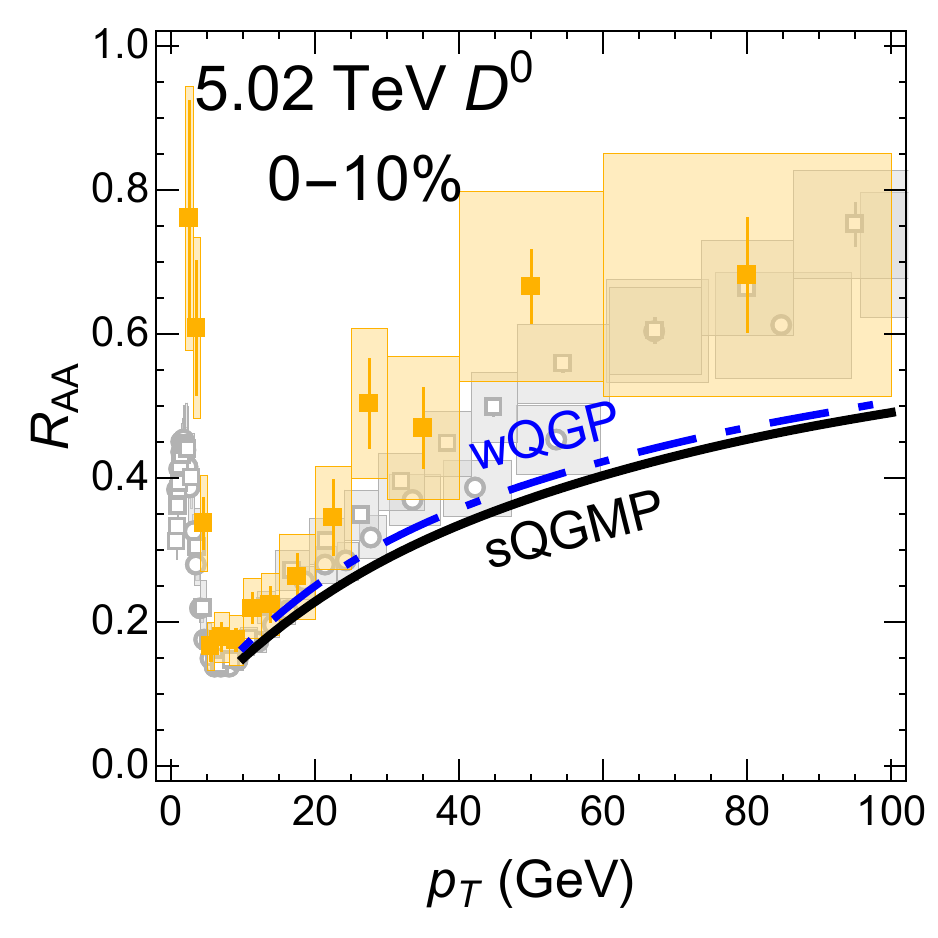}\qquad\qquad
\includegraphics[width=0.3\textwidth]{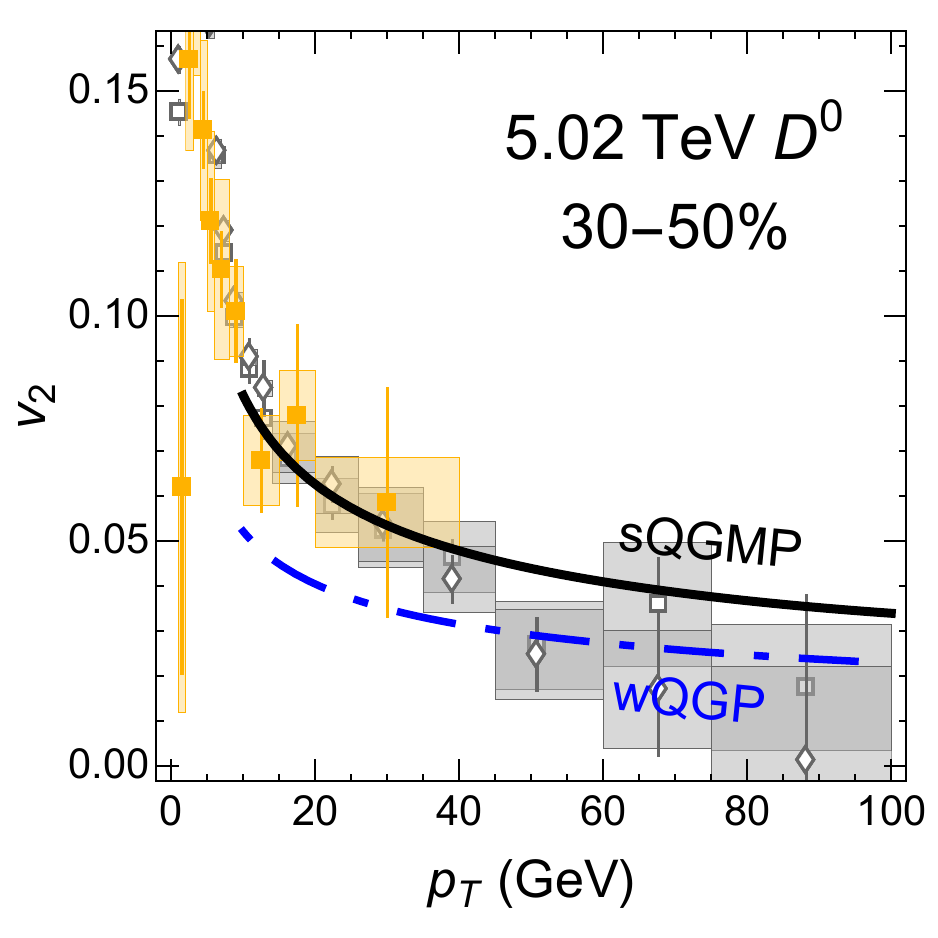}
\caption{(color online) CUJET theoretical expectation of $D$ meson $R_{AA}$ and $v_2$ using: sQGMP $\chi_T^L$-scheme (black solid line), and wQGP/CUJET2 scheme (blue dashed-dotted line). Comparison with preliminary-CMS data (orange solid squares)~\cite{Sirunyan:2017plt,Sirunyan:2017xss} is also shown. Corresponding $R_{AA}$ and $v_2$ data for light hadrons~\cite{ATLAS:2017rmz,Khachatryan:2016odn,Sirunyan:2017pan} are depicted with gray symbols. }
\label{fig.HF}
\end{figure}

With the high statistics of 5.02~TeV Pb-Pb data, we further expect that highly precise jet quenching observables for heavy flavored hadrons, {\it e.g.} $D$ meson, could serve as an independent probe to discriminate sQGMP versus wQGP models.
As shown in Fig.~\ref{fig.HF}, we find the sQGMP and wQGP models predict similar $R_{AA}$, while their significantly different predictions of $v_2$ require experimental data with higher accuracy and higher $p_T$ to provide a decisive distinction.
\clearpage

\section{Comparison with Experimental Data}\label{sec.comparison}
With the systematic $\chi^2$ analysis, we obtained the optimal region of CUJET3 parameters constrained by only light hadron $R_{AA}$ and $v_2$, for central and semi-central collisions.
To provide a critical independent test of the model, we compute CUJET3 results for both light and heavy flavor hadrons, with all centrality ranges up to semi-peripheral collisions, and perform apple-to-apple comparisons with all available experimental data.

Starting from this section, in CUJET3 simulations we employed the $\chi_T^L$-scheme assuming slow quark-libration, while keeping the theoretical uncertainties by taking the parameter region spanned by ($\alpha_c = 0.8$, $c_m = 0.22$) and ($\alpha_c = 1.0$, $c_m = 0.28$), which correspond to upper/lower bounds of $R_{AA}$ and lower/upper bounds of $v_2$, respectively.

\subsection{Light Hadrons}\label{sec.comparison.light}
\begin{figure}[!hbt]\centering
\includegraphics[width=0.95\textwidth]{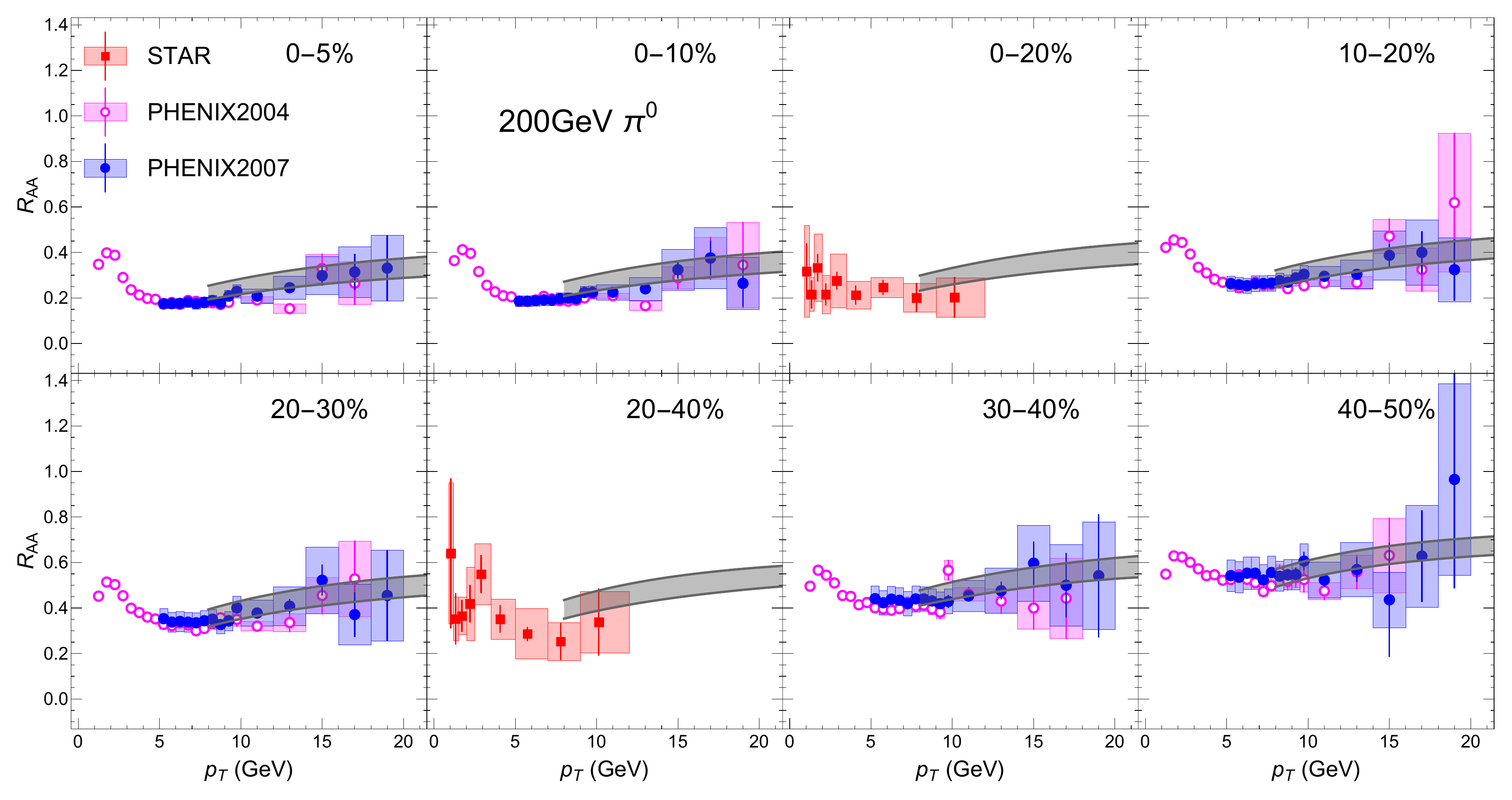}
\caption{(color online) Light hadron $R_{AA}$ for 200~GeV Au-Au collisions in comparison with PHENIX~\cite{Adare:2008qa,Adare:2012wg} and STAR~\cite{Abelev:2009wx} results. Magenta (blue) circles labeled PHENIX2004 (PHENIX2007) correspond to data published in Ref.~\cite{Adare:2008qa} (Ref.~\cite{Adare:2012wg}) analysis of the RHIC 2004 (2007) data set.\label{fig.light_200_RAA}}
~\\~\\
\includegraphics[width=0.99\textwidth]{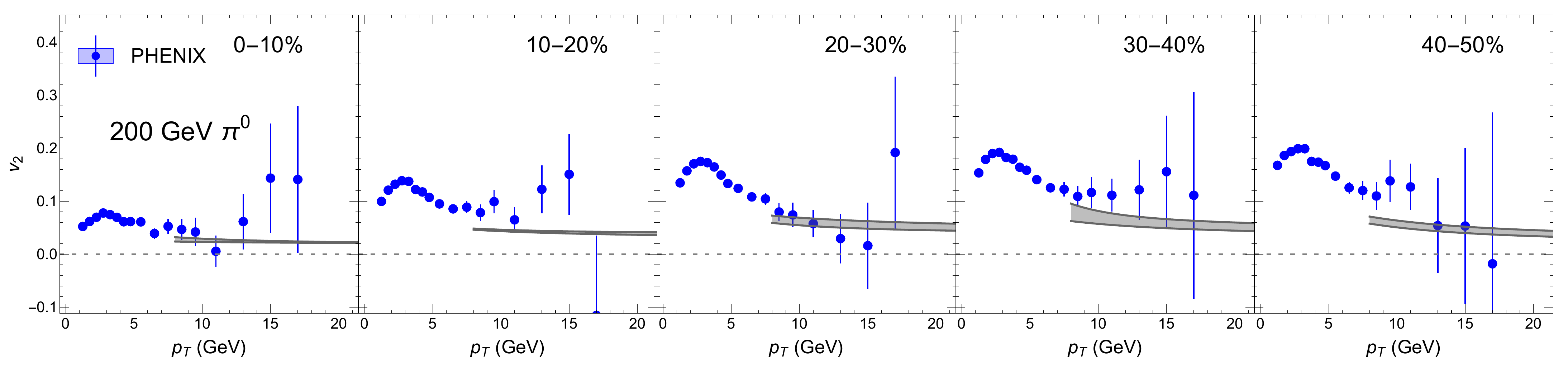}
\caption{(color online) Light hadron $v_{2}$ for 200~GeV Au-Au collisions in comparison with PHENIX data~\cite{Adare:2012wg}.\label{fig.light_200_v2}}
\end{figure}

\begin{figure}[!hbt]\centering
\vspace{-0.5cm}
\includegraphics[width=0.95\textwidth]{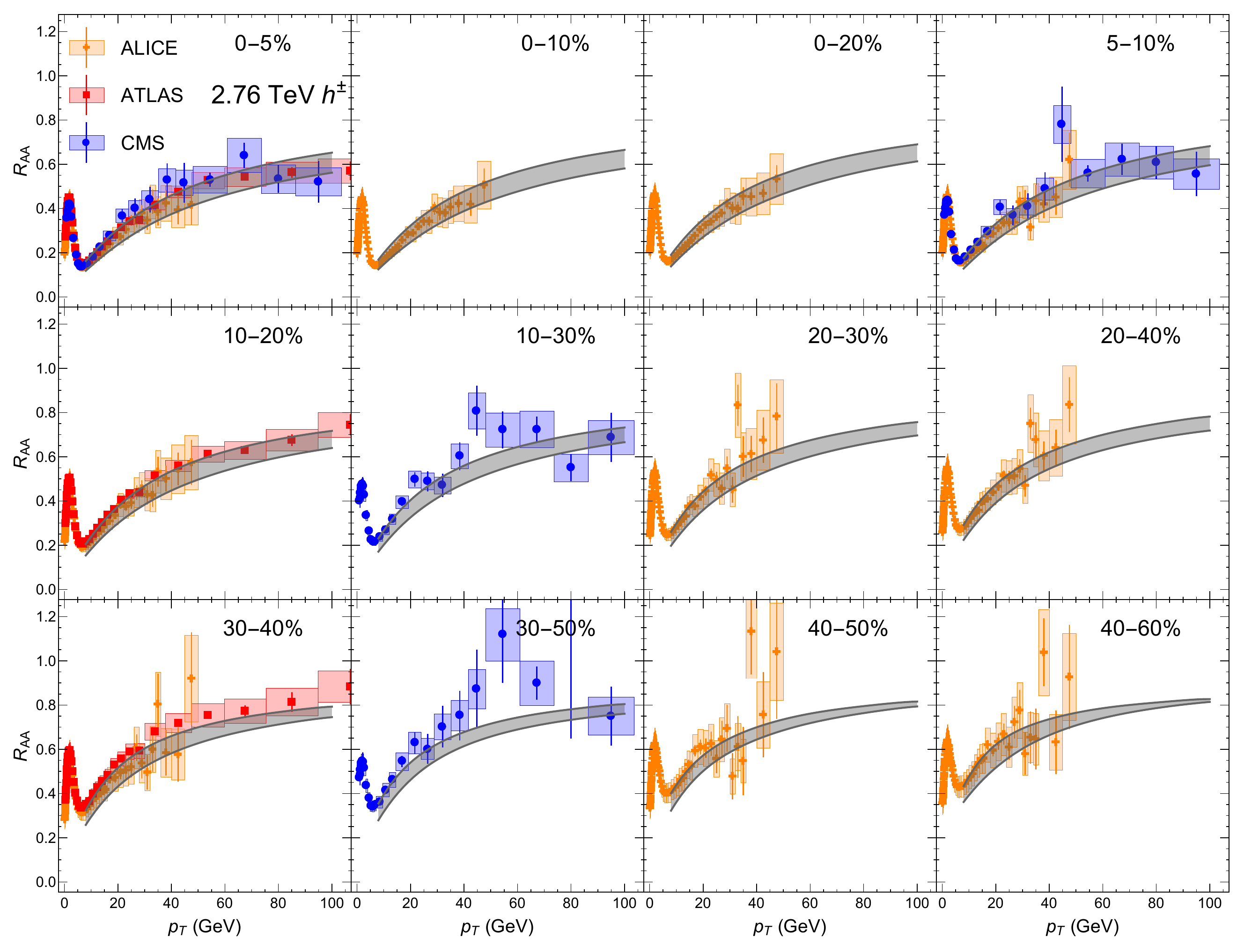}
\caption{(color online) Light hadron $R_{AA}$ for 2.76~TeV Pb-Pb collisions in comparison with ALICE~\cite{Abelev:2012hxa}, ATLAS~\cite{Aad:2015wga} and CMS~\cite{CMS:2012aa} results.\label{fig.light_2760_RAA}}
~\\
\vspace{-0.5cm}
\includegraphics[width=0.95\textwidth]{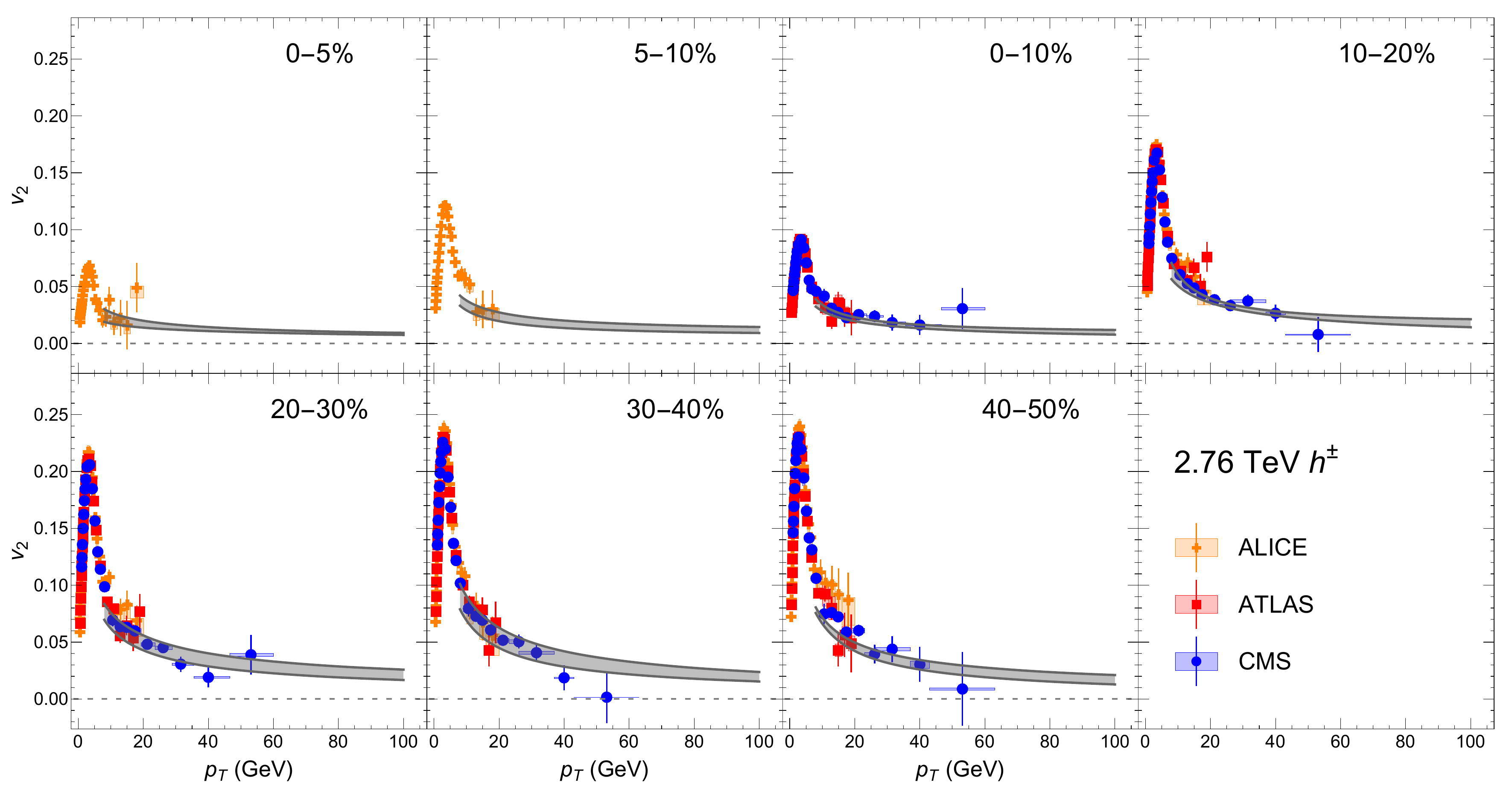}
\caption{(color online) Light hadron $v_{2}$ for 2.76~TeV Pb-Pb collisions in comparison with ALICE~\cite{Abelev:2012di}, ATLAS~\cite{ATLAS:2011ah} and CMS~\cite{Chatrchyan:2012xq} results.\label{fig.light_2760_v2}}
\end{figure}

\begin{figure}[!hbt]\centering
\includegraphics[width=0.95\textwidth]{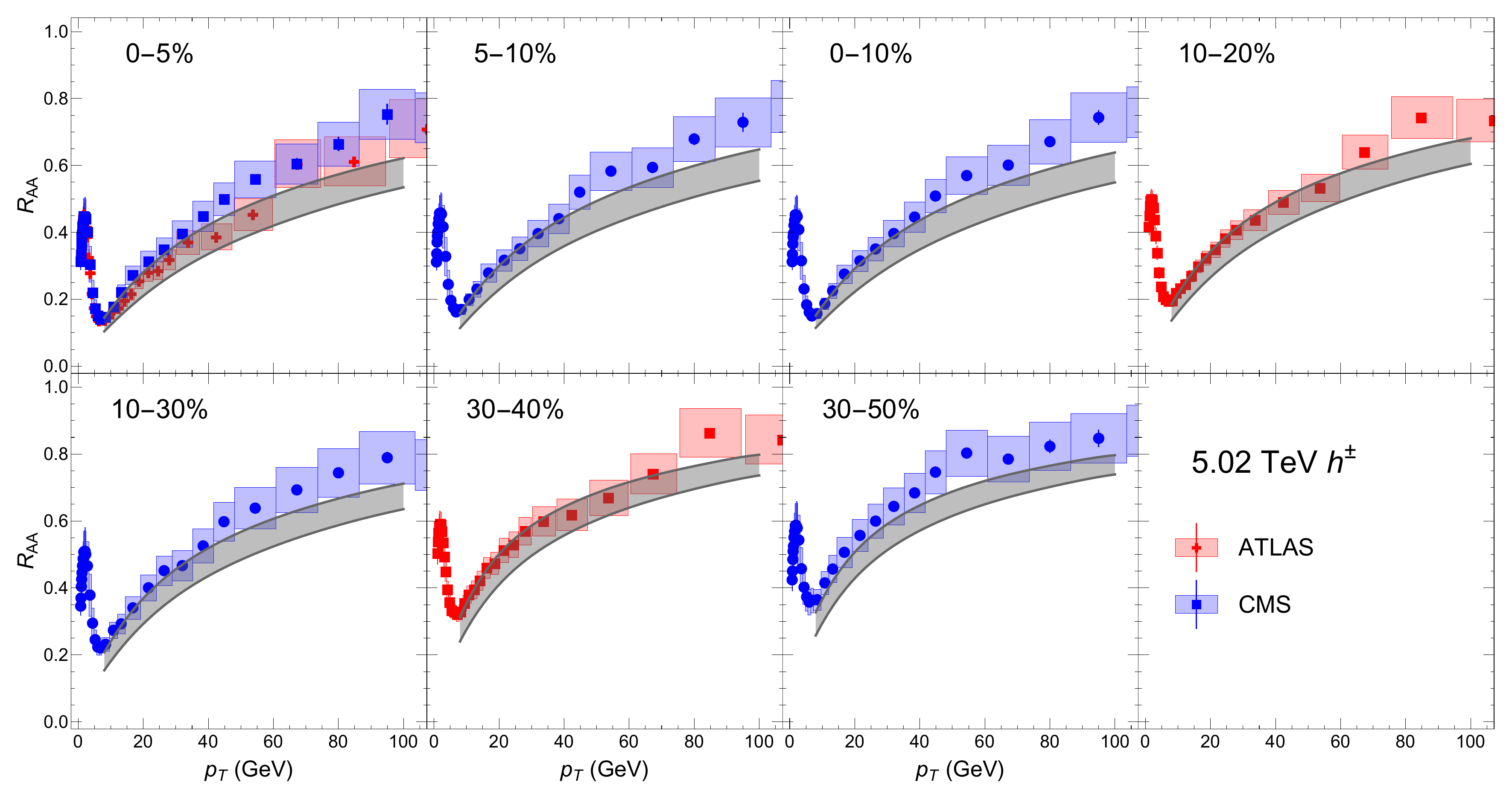}
\caption{(color online) Light hadron $R_{AA}$ for 5.02~TeV Pb-Pb collisions in comparison with ATLAS~\cite{ATLAS:2017rmz} and CMS~\cite{Khachatryan:2016odn} results.\label{fig.light_5020_RAA}}
~\\~\\
\includegraphics[width=0.7\textwidth]{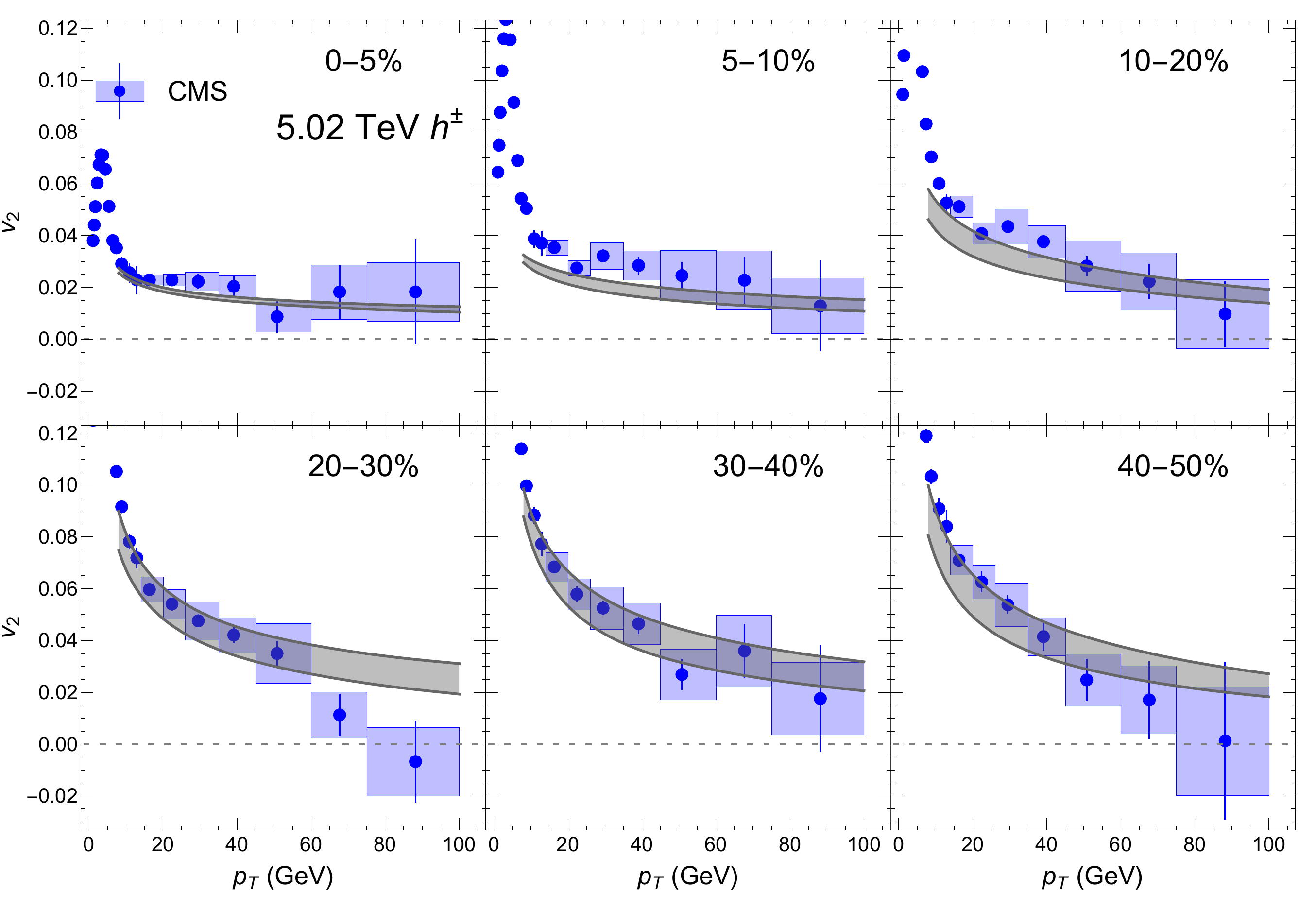}
\caption{(color online) Light hadron $v_{2}$ for 5.02~TeV Pb-Pb collisions in comparison with CMS data~\cite{Sirunyan:2017pan}.\label{fig.light_5020_v2}}
\end{figure}

First of all, in Figures~\ref{fig.light_200_RAA}-\ref{fig.light_5020_v2}, we compare CUJET3.1 results for light hadron $R_{AA}$ and $v_2$, with all available data:
PHENIX~\cite{Adare:2008qa,Adare:2012wg} and STAR~\cite{Abelev:2009wx} measurements for 200~GeV Au-Au collisions;
ALICE~\cite{Abelev:2012hxa,Abelev:2012di}, ATLAS~\cite{Aad:2015wga,ATLAS:2011ah} and CMS~\cite{CMS:2012aa,Chatrchyan:2012xq} results for 2.76~TeV Pb-Pb collisions;
and ATLAS~\cite{ATLAS:2017rmz} and CMS~\cite{Khachatryan:2016odn,Sirunyan:2017pan} data for 5.02~TeV Pb-Pb collisions.
One can clearly see the excellent agreement for all centrality ranges at all mentioned collision energies. 
In particular, it is worth to emphasize that after the aforementioned correction, the current CUJET3.1 simulation framework is able to correctly reproduce the $p_T$ and centrality dependence of both $R_{AA}$ and $v_2$.

\begin{table}[!hbt]\centering
\begin{tabular}{c|p{2mm}cp{2mm}cp{2mm}|p{2mm}cp{2mm}cp{2mm}|p{2mm}cp{2mm}cp{2mm}c}
\hline\hline
		& \multicolumn{5}{c}{200~GeV} & \multicolumn{5}{|c|}{2.76~TeV} & \multicolumn{5}{c}{5.02~TeV} &\\
\cline{2-17}
		&& 0\%--5\% && 40\%--50\% &&& 0\%--5\% && 40\%--50\% &&& 0\%--5\% && 40\%--50\%  &  \\
\hline
$T_{\mathrm{ini,center}}$ (MeV)	&& 358 && 294 &&& 465 && 366 &&& 506 && 397 \\
$\epsilon_{2,\mathrm{ini}}$		&& 0.07  && 0.44 &&& 0.07 && 0.46  &&& 0.07 && 0.45  \\
$\tau_{\mathrm{hydro}}$ (fm/c) 		&& 9.4 && 5.2 &&& 11.4 && 6.3 &&& 11.8 && 6.7  \\
\hline
\hline
\end{tabular}
\caption{Comparison of the initial central temperature $T_{\mathrm{ini,center}}$, initial ellipticity $\epsilon_{2,\mathrm{ini}}$, and life-time $\tau_{\mathrm{hydro}}$
in different collision conditions.
The initial ellipticity is defined with respect to entropy density $s$ at hydro starting time $\tau=0.6$~fm, 
$\epsilon_{2,\mathrm{ini}} \equiv -{[\int s\; \rho^2 \cos(2\phi)\; dx dy]}/{[\int s\;\rho^2\; dx dy]}.$
\label{tab.bulkinfo}}
\end{table}

We note that such a comprehensive data set covers a rich diversity of geometrical and thermal profiles of the QCD Plasma.
In different centrality bins at various colliding energies, the bulk backgrounds are significantly distinctive in lifetime, size, ellipticity and temperature, 
and consequently, the path length of the jets, either direction averaging or depending, varies in a wide range.
In Tab.~\ref{tab.bulkinfo}, we show the quantitative comparison of the initial central temperature $T_{\mathrm{ini,center}}$, initial ellipticity $\epsilon_{2,\mathrm{ini}}$, and life time $\tau_{\mathrm{hydro}}$.
The temperature, as well as the life-time of such systems vary by a factor of $\sim2$, while the geometries change from nearly symmetric to those with an ellipticity $\sim0.4$.
The success in explaining $R_{AA}$ and $v_2$ from central to semi-peripheral data, at beam energies from 0.2~TeV to 5.02~TeV, indicates the success of the temperature and path dependence of the CUJET3 energy loss model.

\clearpage

\subsection{Heavy Flavor Measurements}
\begin{figure}[!hbt]\centering
\includegraphics[width=0.7\textwidth]{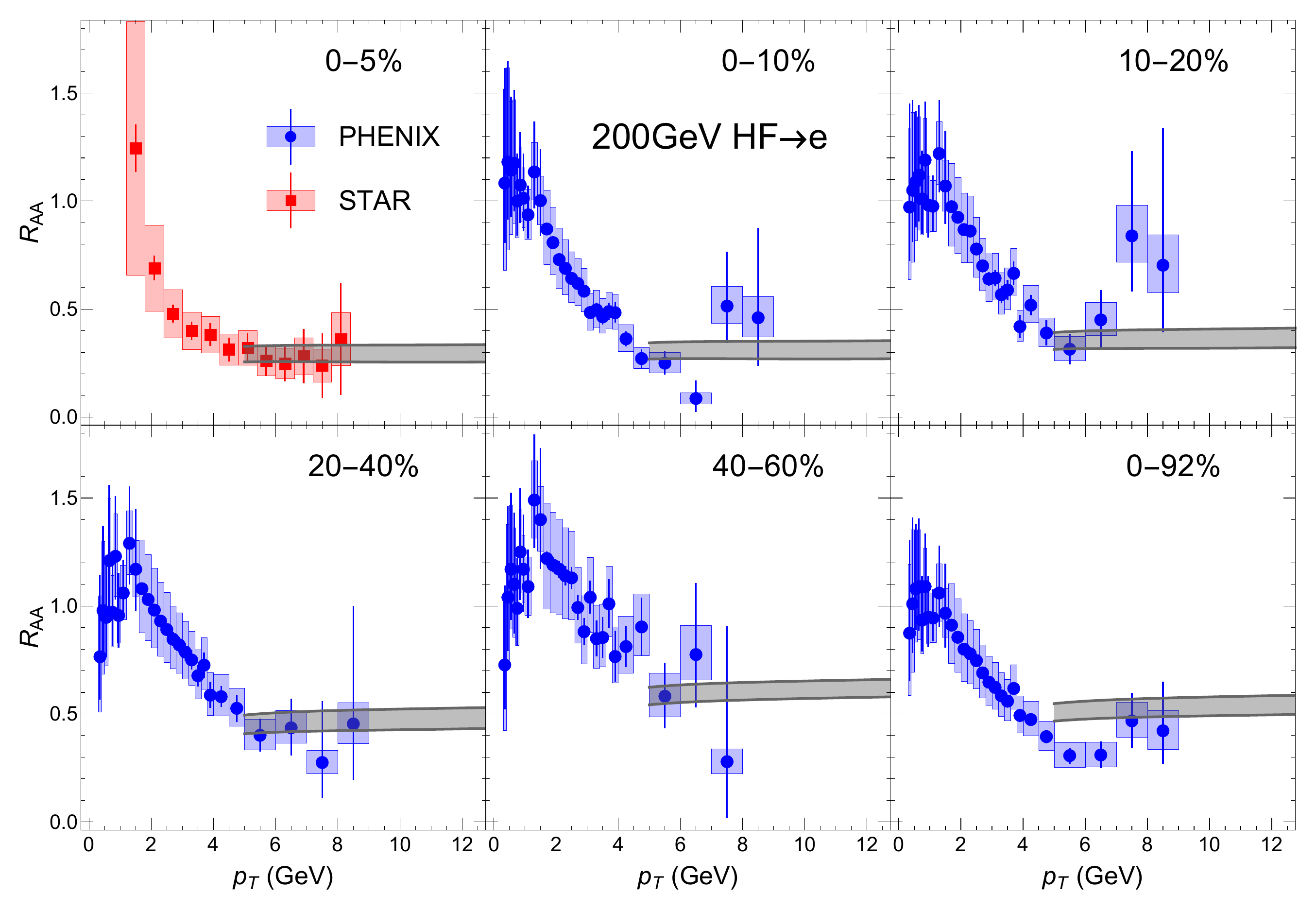}
\caption{(color online) Heavy flavor decayed electron $R_{AA}$ for 200~GeV Au-Au collisions in comparison with PHENIX~\cite{Adare:2010de} and STAR~\cite{Abelev:2006db} results.
\label{fig.heavy_200_eRAA}}
~\\
\includegraphics[width=0.497\textwidth]{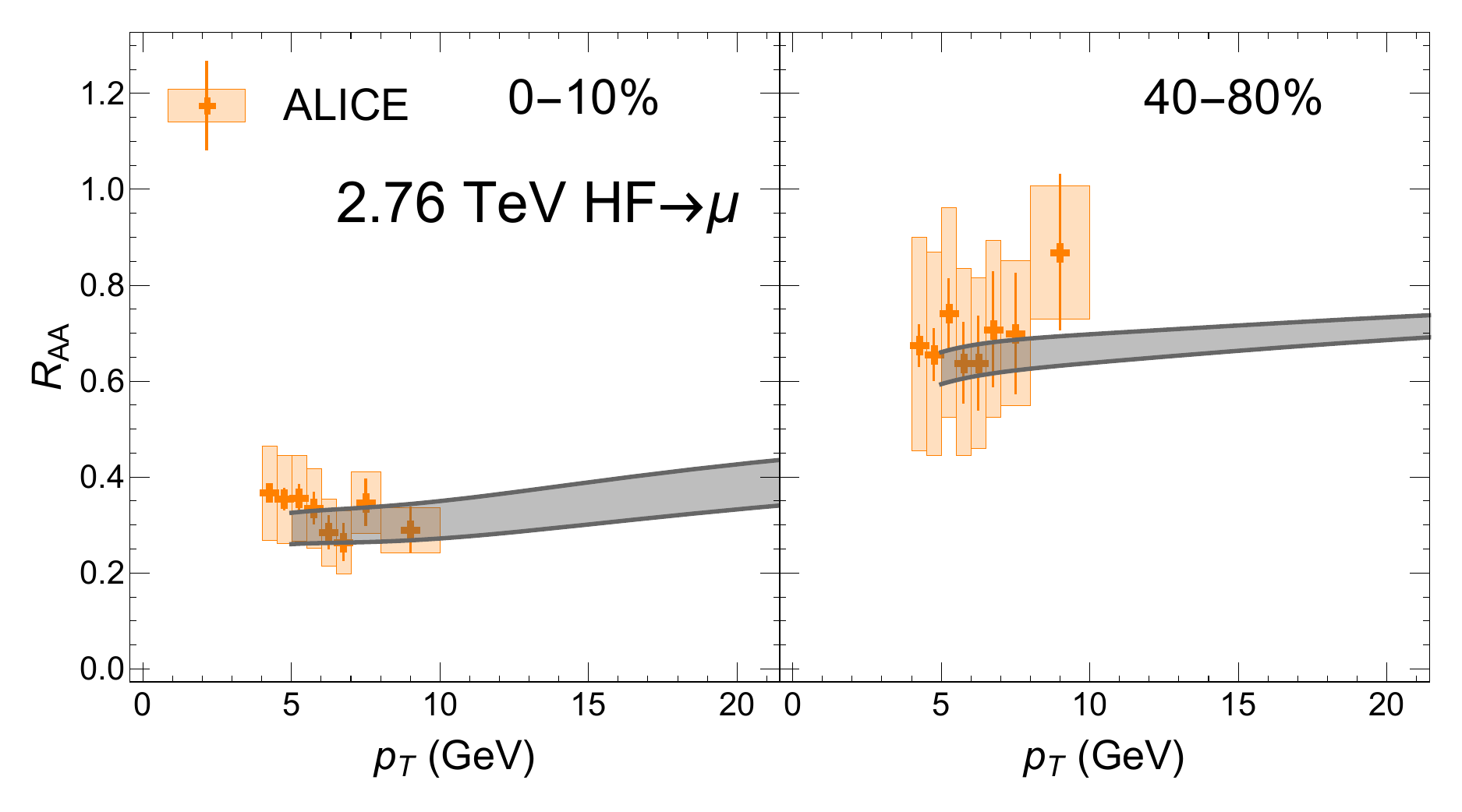}
\caption{(color online) Heavy flavor decayed muon $R_{AA}$ for 2.76~TeV Pb-Pb collisions in comparison with ALICE data~\cite{Abelev:2012qh}.
\label{fig.heavy_2760_muRAA}}
\end{figure}

\begin{figure}[!hbt]\centering
\includegraphics[width=0.99\textwidth]{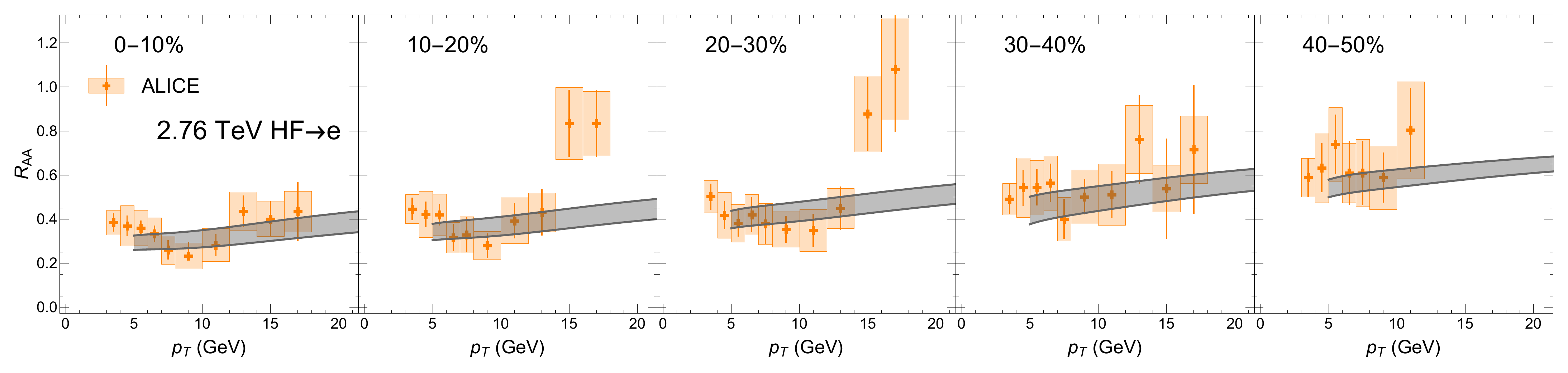}
\caption{(color online) Heavy flavor decayed electron $R_{AA}$ for 2.76~TeV Pb-Pb collisions in comparison with ALICE data~\cite{Adam:2016khe}.
\label{fig.heavy_2760_eRAA}}
~\\
\includegraphics[width=0.7\textwidth]{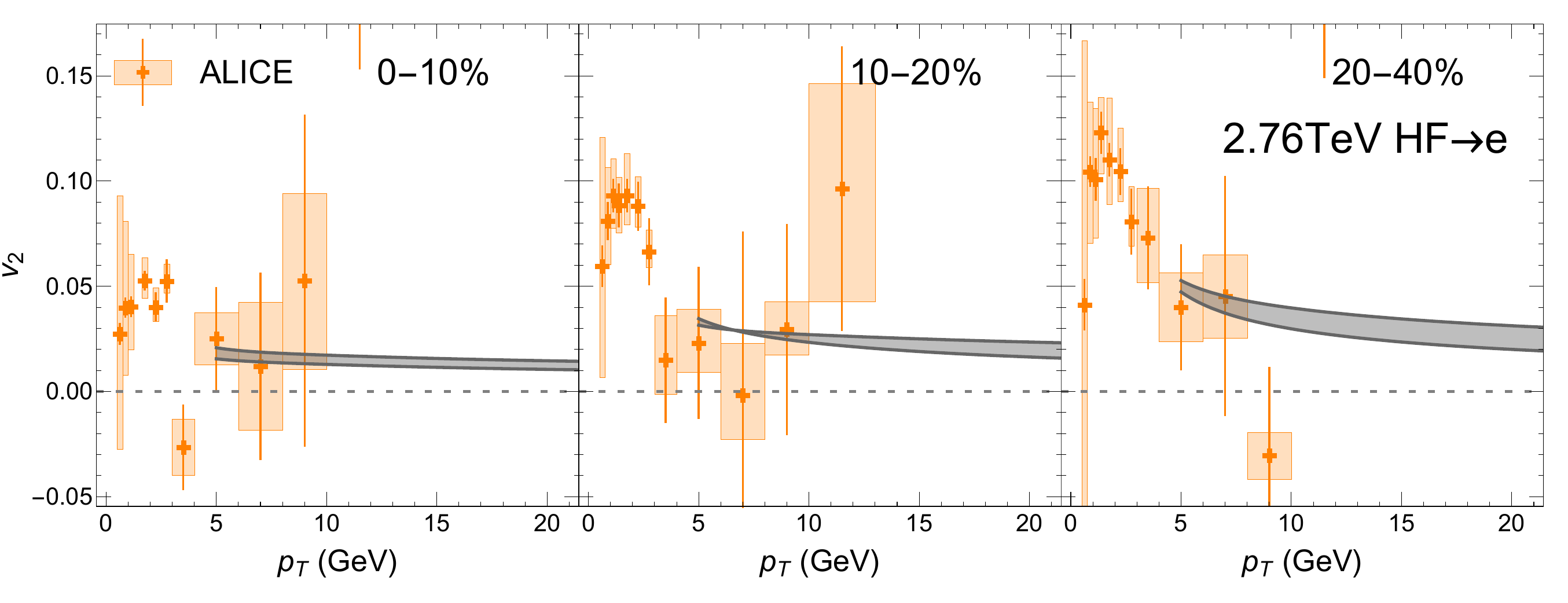}
\caption{(color online) Heavy flavor decayed electron $v_{2}$ for 2.76~TeV Pb-Pb collisions in comparison with ALICE data~\cite{Adam:2016ssk}.
\label{fig.heavy_2760_ev2}} 
~\\
\includegraphics[width=0.7\textwidth]{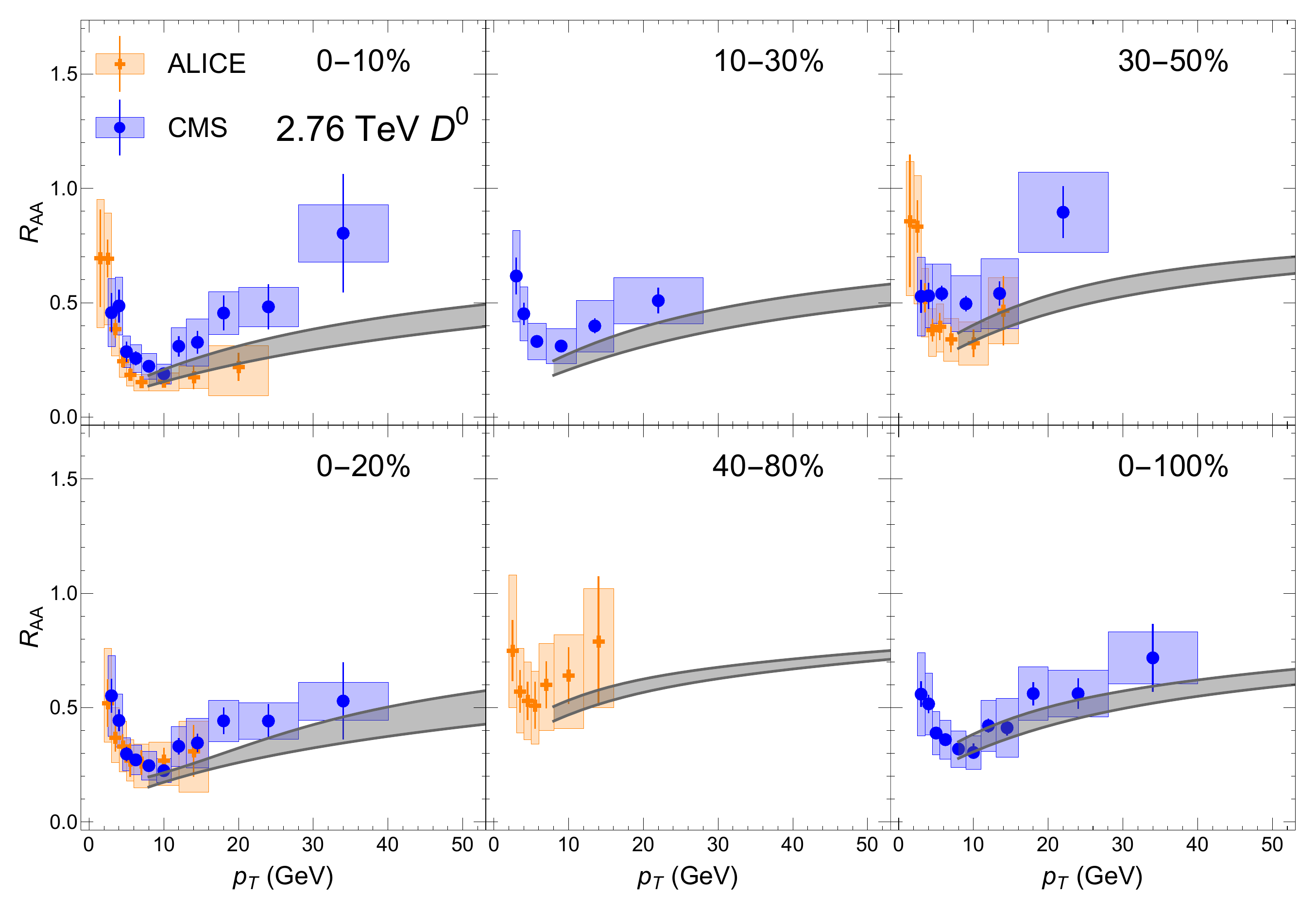}
\caption{(color online) Prompt $D$ meson $R_{AA}$ for 2.76~TeV Pb-Pb collisions in comparison with ALICE~\cite{ALICE:2012ab} and preliminary CMS~\cite{CMS:2015hca} results.
\label{fig.heavy_2760_DRAA}}
\end{figure}

\begin{figure}[!hbt]\centering
\vspace{-0.5cm}
\includegraphics[width=0.724\textwidth]{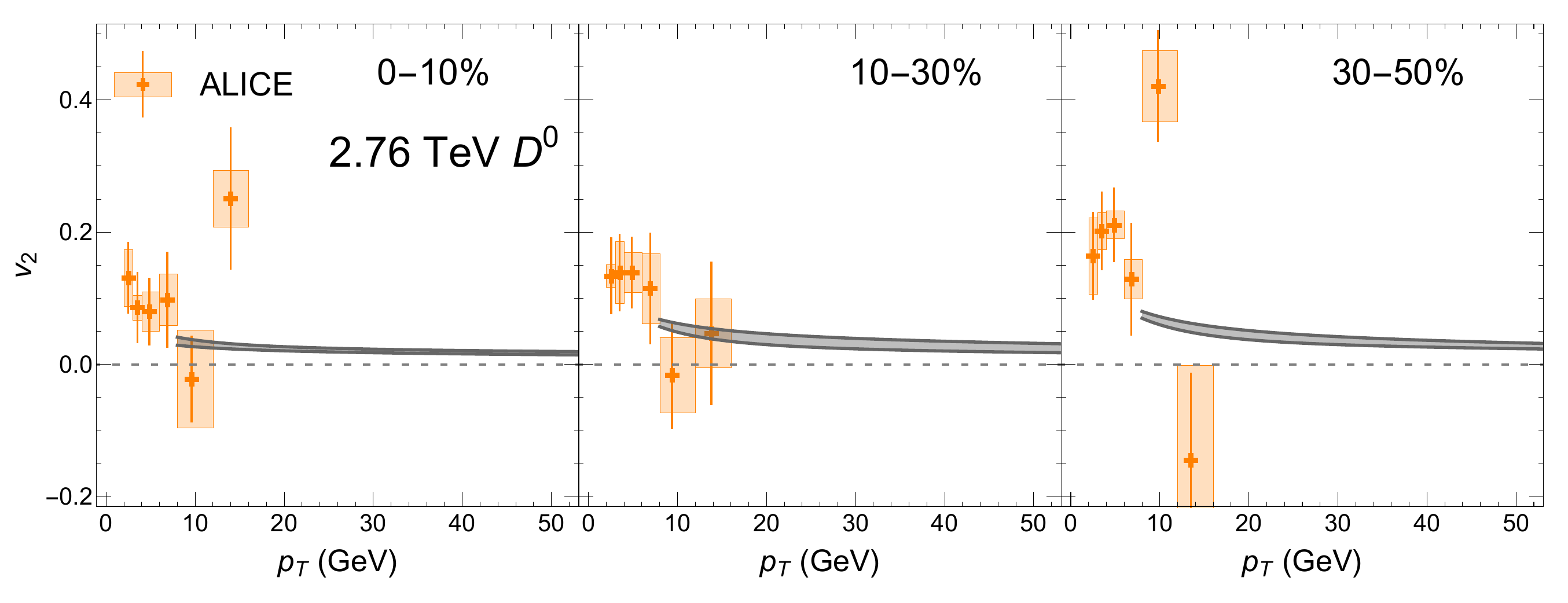}
\caption{(color online) $D$ meson $v_{2}$ for 2.76~TeV Pb-Pb collisions in comparison with ALICE data~\cite{Abelev:2014ipa}.
\label{fig.heavy_2760_Dv2}}
\vspace{0.2cm}
\includegraphics[width=0.95\textwidth]{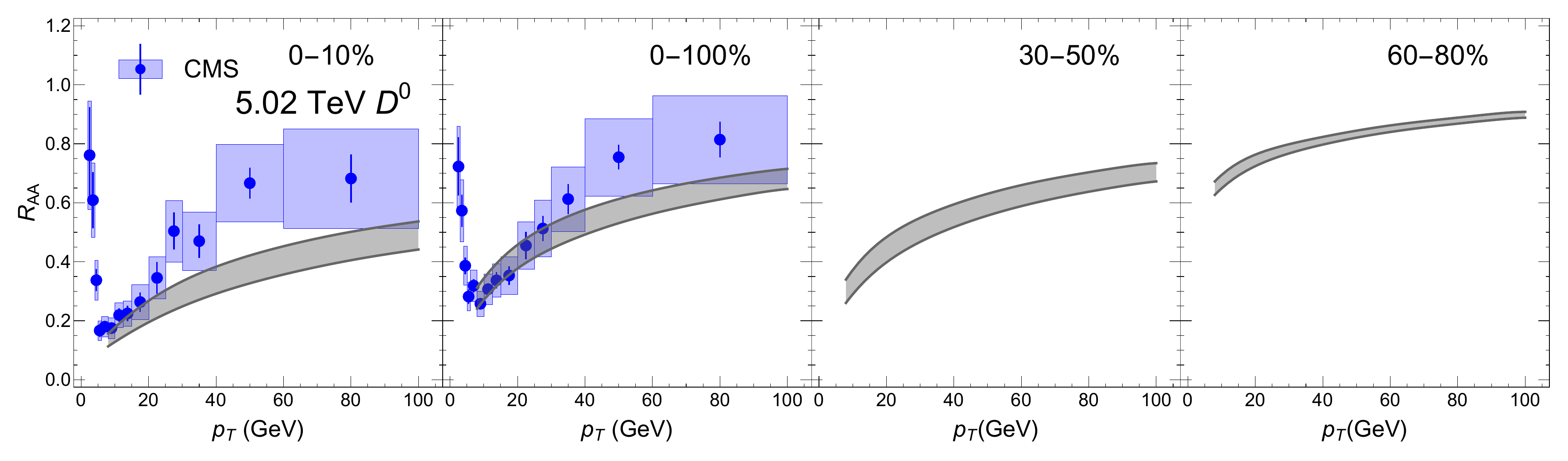}
\caption{(color online) Prompt $D$ meson $R_{AA}$ for 5.02~TeV Pb-Pb collisions in comparison with preliminary CMS data~\cite{Sirunyan:2017xss}.
\label{fig.heavy_5020_DRAA}}
\vspace{0.2cm}
\includegraphics[width=0.95\textwidth]{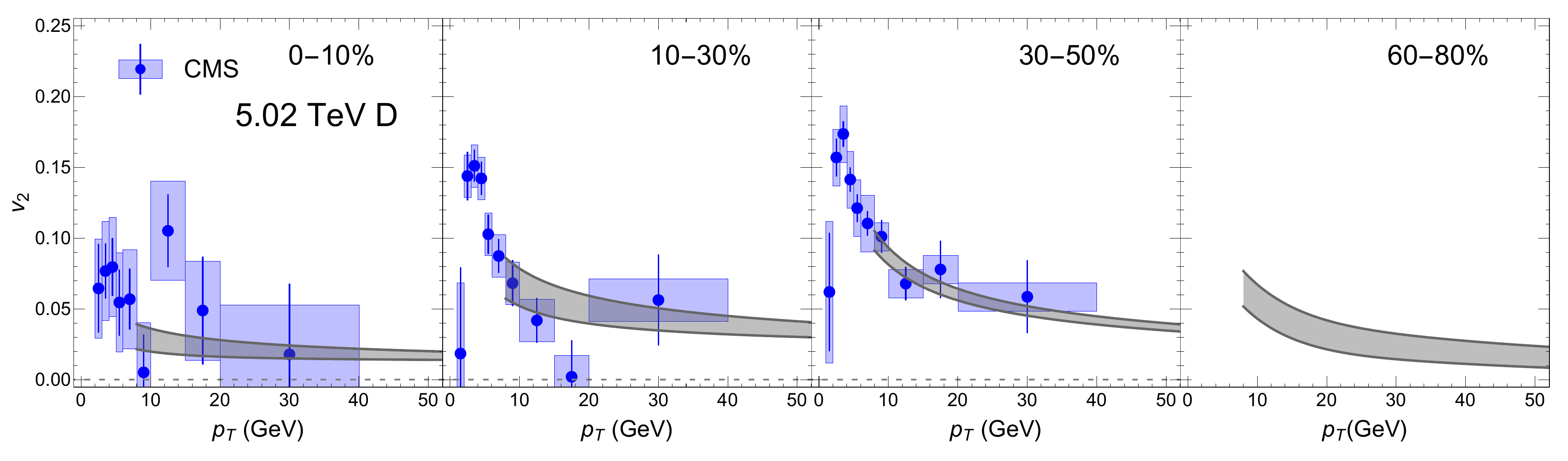}
\caption{(color online) Prompt $D$ meson $v_{2}$ for 5.02~TeV Pb-Pb collisions in comparison with preliminary CMS data~\cite{Sirunyan:2017plt}.
\label{fig.heavy_5020_HFv2}}
\vspace{0.2cm}
\includegraphics[width=0.272\textwidth]{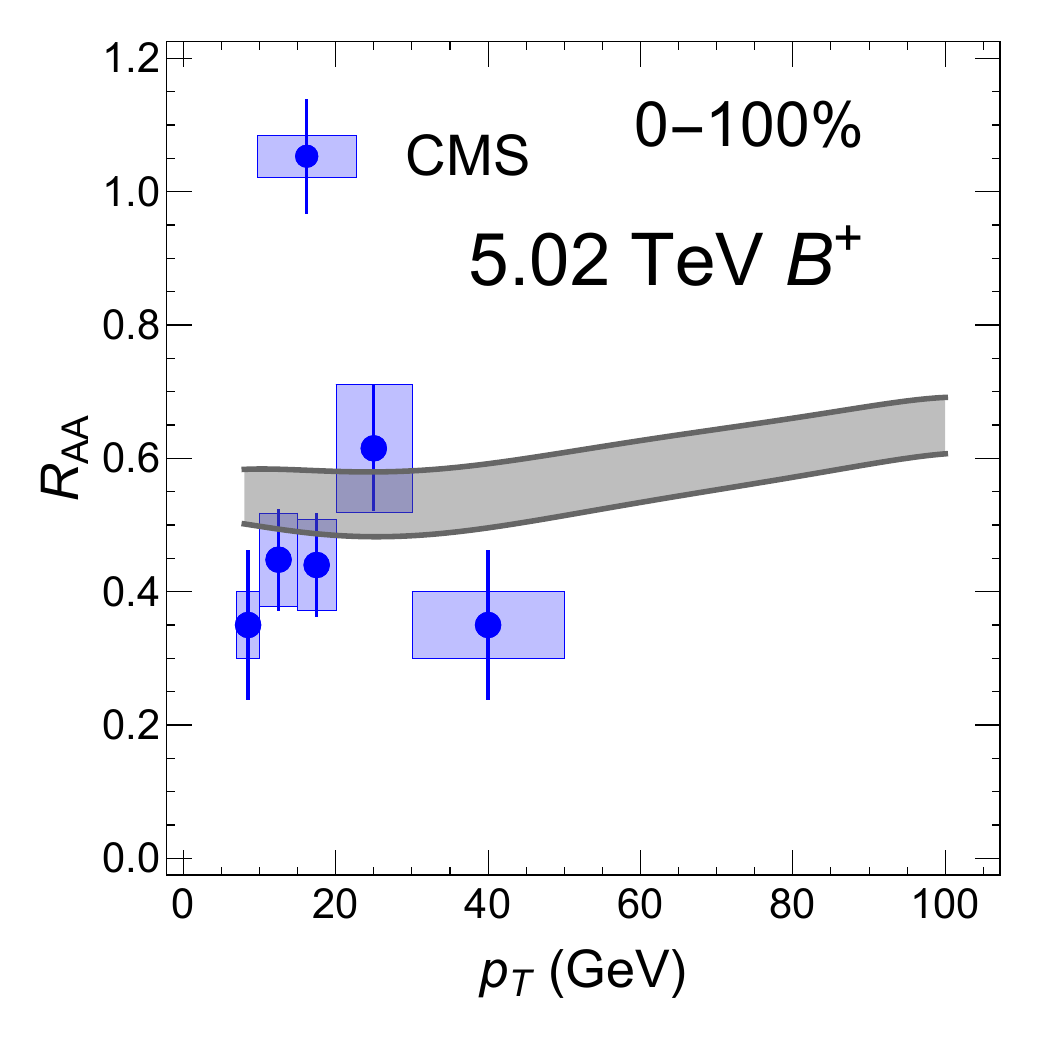}
\caption{(color online) Prompt $B$ meson $R_{AA}$ for 5.02~TeV Pb-Pb collisions in comparison with CMS data~\cite{Sirunyan:2017oug}.
\label{fig.heavy_5020_BRAA}}
\end{figure}

Having successfully described high-$p_T$ $R_{AA}$ and $v_2$ data for light hadrons, we now perform further {\em independent tests} of the energy-loss mechanism using heavy flavor data~\cite{Rapp:2018qla}.
In Figures~\ref{fig.heavy_200_eRAA}-\ref{fig.heavy_5020_BRAA}, we compare CUJET3 results for the energy-loss observables of prompt $D$ \& $B$ mesons as well as electrons or muons from heavy flavor decay, with all available data:
PHENIX~\cite{Adare:2010de}, STAR~\cite{Abelev:2006db} measurements for 200~GeV Au-Au collisions;
ALICE~\cite{ALICE:2012ab,Abelev:2014ipa,Adam:2016khe,Adam:2016ssk,Abelev:2012qh}, CMS~\cite{CMS:2015hca} data for 2.76~TeV Pb-Pb collisions;
and finally CMS results~\cite{Sirunyan:2017xss,Sirunyan:2017oug,Sirunyan:2017plt} for 5.02~TeV Pb-Pb collisions.
A very good agreement between model and data is found, which validate a successful and unified description of CUJET3 for both light and heavy flavor jet energy loss observables.


\clearpage

\section{CUJET3 Predictions for Other Experimental Observables}\label{sec.prediction}
In the above section we perform a successful test of the CUJET3 framework, which provides a united description for comprehensive sets of experimental data, from average suppression to azimuthal anisotropy, from light flavor to heavy flavor observables, with beam energies from $200$~GeV to $5.02$~TeV, and from central to semi-peripheral collisions. 
With the new colliding system or new experimental observables, we expect more stringent tests to help further constrain the CUJET3 energy loss model.
In this section, we show the CUJET3 prediction for ongoing experimental analysis, including jet quenching observables in $^{129}_{\;\;54}$Xe + $^{129}_{\;\;54}$Xe collisions at 5.44~TeV and more heavy flavor signals in 5.02~TeV Pb-Pb collisions.

\subsection{Light Hadron $R_{AA}$ in 5.44~TeV $^{129}_{\;\;54}$Xe-$^{129}_{\;\;54}$Xe Collisions}
\begin{figure}[!hbt]\centering
\includegraphics[width=0.95\textwidth]{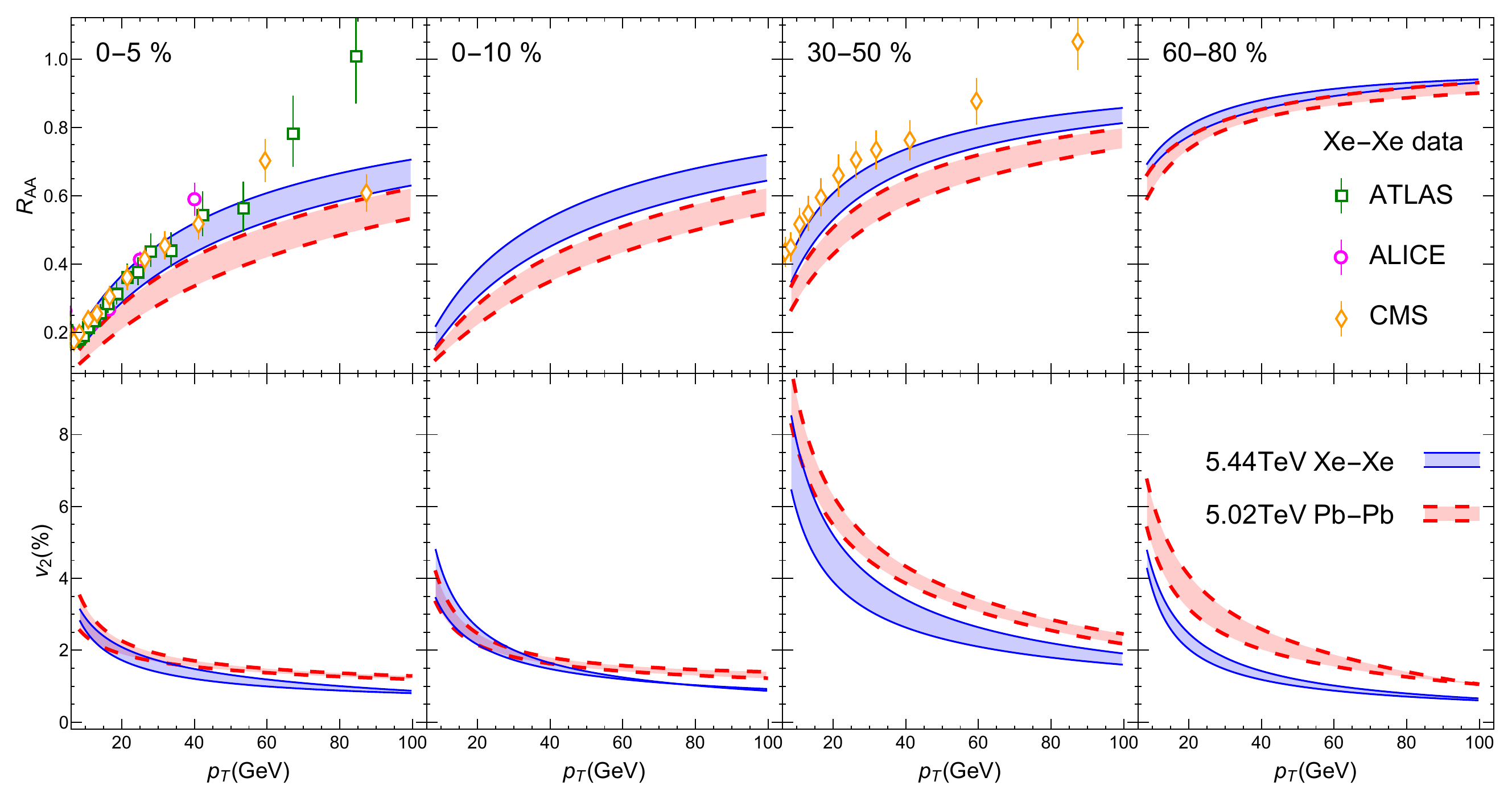}
\caption{(color online) Light hadron $R_{AA}$ and $v_2$ for $5.44$~TeV Xe-Xe collisions (blue bands) and $5.02$~TeV Pb-Pb collisions (red dashed curves). 
Preliminary experimental data~\cite{Acharya:2018eaq,ATLAS:2018vmo,Sirunyan:2018eqi} are also shown.}
\label{fig.pre_5440XeXe}
\end{figure}
Recently the LHC ran collisions with a new species of nuclei, colliding xenon with 129 nucleons ($^{129}_{\;\;54}$Xe), at a beam energy of $\sqrt{s_{NN}}=5.44$~TeV.
In Xe-Xe collisions, the hot medium created is expected to be a bit cooler and shorter lived when compared with the one created in 5.02~TeV Pb-Pb collisions.
Given the similar beam energy, it's expected that the difference between observables from these two colliding system provide valuable information on the nature of the QGP, especially on how the hot medium interacts with high energy jets.

In Fig.~\ref{fig.pre_5440XeXe} we show the light hadron $R_{AA}$ and $v_2$ for both systems.
Higher $R_{AA}$ and lower $v_{2}$ in $5.44$~TeV Xe-Xe collisions (blue bands) are produced, compared with those in $5.02$~TeV Pb-Pb collisions (red dashed curves).
This indicates that the high-$p_T$ light hadrons produced in the former system are less suppressed than those produced in latter,
exhibiting the sensitivity of the jet-quenching observables to the system size and density: 
when comparing to those created in Pb-Pb collisions, jets created in Xe-Xe collisions travel a shorter path in the hot medium and interact with less dense matter, hence losing less energy.
With this new colliding system, we are able to further test the path length dependence of the CUJET3 jet energy loss model.
Such predictions were made before the experimental measurements reported at the Quark Matter 2018 conference. Our predictions are in good agreement with the recently released preliminary data for charged hadron $R_{AA}$ from the ALICE~\cite{Acharya:2018eaq}, ATLAS~\cite{ATLAS:2018vmo}, and CMS~\cite{Sirunyan:2018eqi} collaborations (as shown in Fig.~\ref{fig.pre_5440XeXe}). See also Ref.~\cite{Sirunyan:2018eqi} for a detailed data-model comparison. 

\subsection{$B$-decayed $D$ Meson $R_{AA}$ in $5.02$~TeV Pb-Pb Collisions}
\begin{figure}[!hbt]\centering
\includegraphics[width=0.724\textwidth]{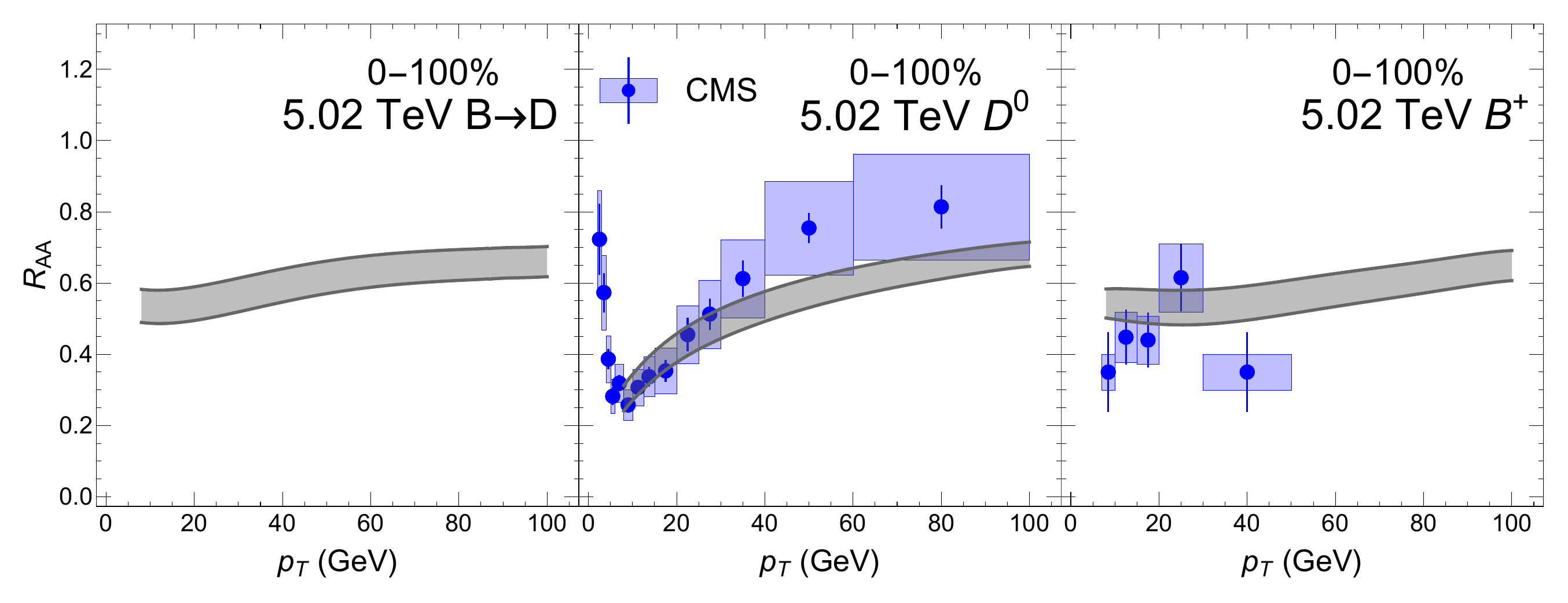}
\caption{(color online) $R_{AA}$ for $D$ meson from $B$-decay (left), prompt $D$ meson (middle), and $B$ meson (right) in minimal-bias $5.02$~TeV Pb-Pb collisions.}
\label{fig.pre_5020_B2D}
\end{figure}
Another new experimental measurement is the $B$-decayed $D$ meson $R_{AA}$ in $5.02$~TeV Pb-Pb collisions.
As shown in Fig.~\ref{fig.pre_5020_B2D}, the $R_{AA}$ of $B$-decay $D$ meson (left panel) has similar $p_T$-dependence as that of $B$ mesons (right panel), and both of them are less suppressed than the prompt $D$ meson (middle panel), especially in the region with lower momentum ($p_T<20$~GeV).
We expect that the future precise measurement of $B$-decay $D$ meson $R_{AA}$ will provide observation of the ``dead cone'' effect, which suppresses the radiational energy loss of bottom jets.

\subsection{High-$p_T$ $D$ mesons in 200~GeV Au-Au collisions}
\begin{figure}[!hbt]\centering
\includegraphics[width=0.724\textwidth]{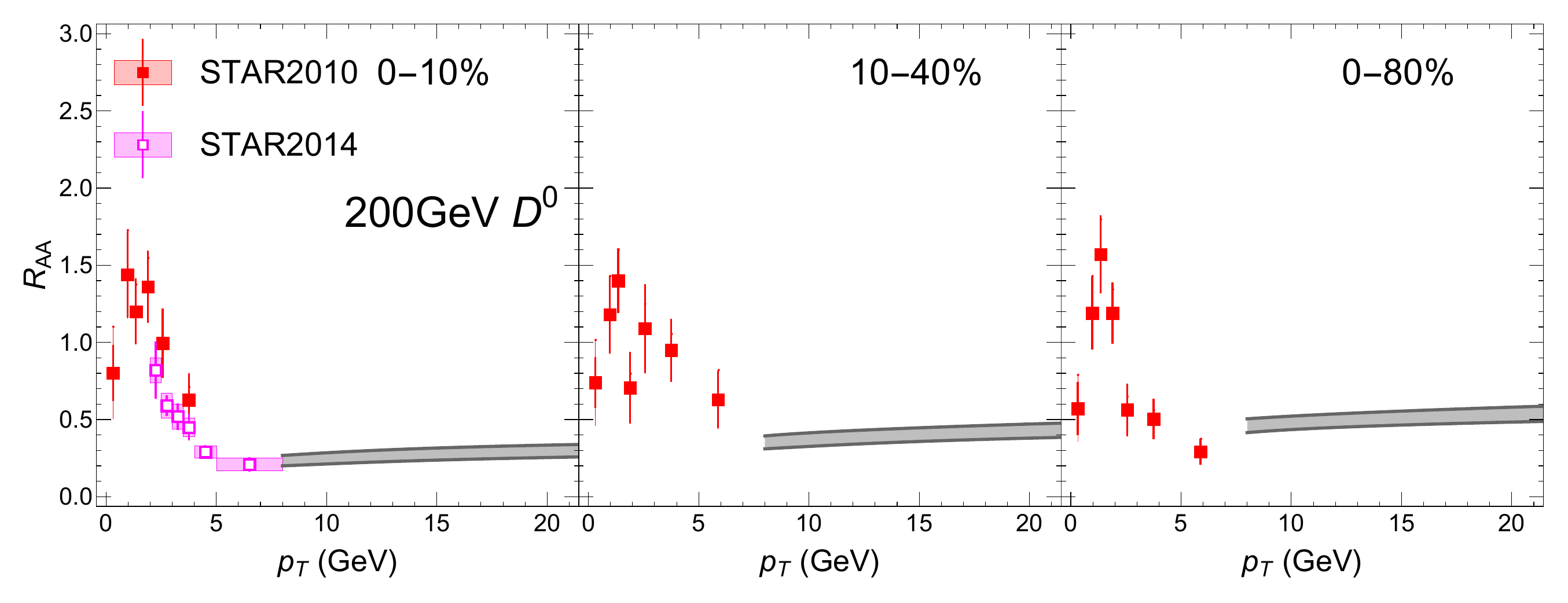}
\includegraphics[width=0.497\textwidth]{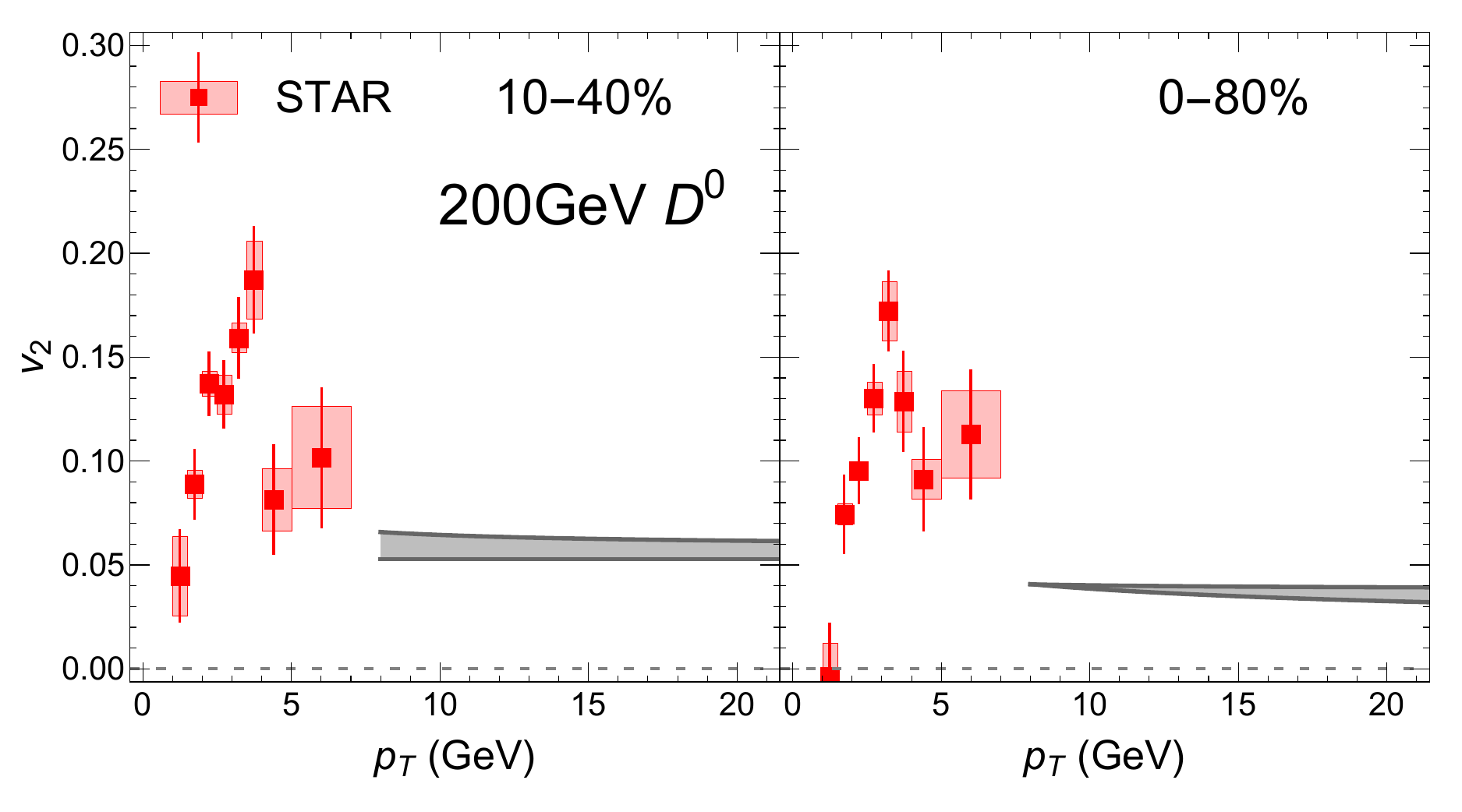}
\caption{ $D$ meson $R_{AA}$ and $v_2$ in 200~GeV Au-Au collisions. STAR data~\cite{Adamczyk:2014uip,Adamczyk:2017xur,Lomnitz:2017mue} for lower $p_T$ range are also shown. Red (magenta) symbols labeled STAR2010 (STAR2014) correspond to data published in Ref.~\cite{Adamczyk:2014uip} (Ref.~\cite{Lomnitz:2017mue}) analysis of the RHIC 2010/11 (2014) data set.}
\label{fig.pre_200_D}
\end{figure}

Recently, the STAR Collaboration at RHIC installed the Heavy Flavor Tracker, which allows high precision measurements of open heavy flavor hadrons.
Early results of azimuthal anisotropy for lower $p_T$ $D$ mesons has shown interesting properties of the low energy charm quarks~\cite{Adamczyk:2017xur}.
With the CUJET3 predictions for $D$ meson's $R_{AA}$ and $v_2$ shown in Fig.~\ref{fig.pre_200_D}, precise measurements of high $p_T$ $D$ meson jet quenching observable could enable the direct comparison with heavy flavor data, and further test the consistency of the HF sector of CUJET3 energy loss for different beam energies.

\subsection{Heavy Flavor Decayed Leptons in 5.02~TeV Pb-Pb Collisions}
\begin{figure}[!hbt]\centering
\includegraphics[width=0.99\textwidth]{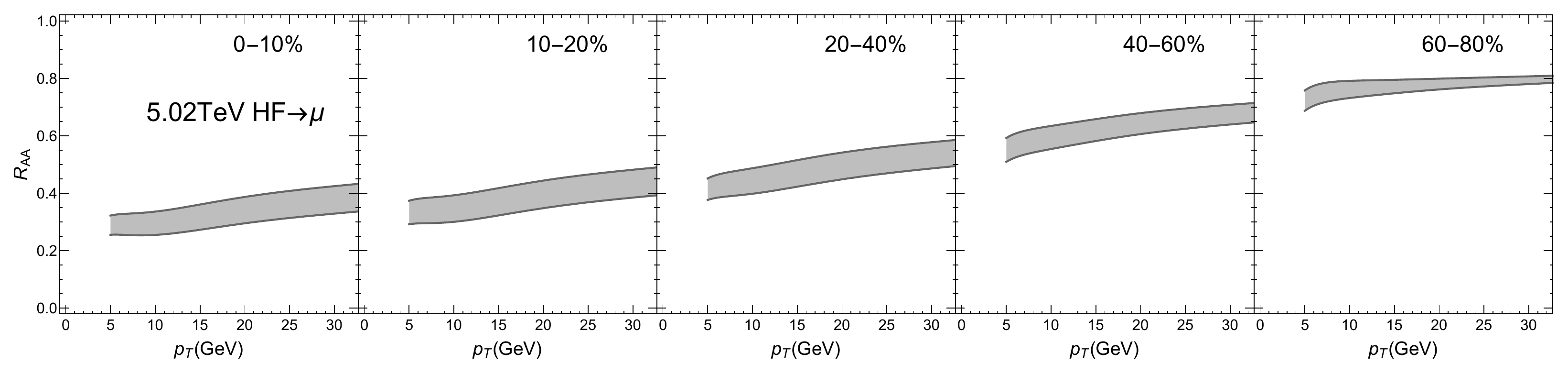}
\caption{$R_{AA}$ for heavy flavor decayed muon in 5.02~TeV Pb-Pb collisions.}
\label{fig.pre_5020_mu}
\end{figure}

\begin{figure}[!hbt]\centering
\includegraphics[width=0.272\textwidth]{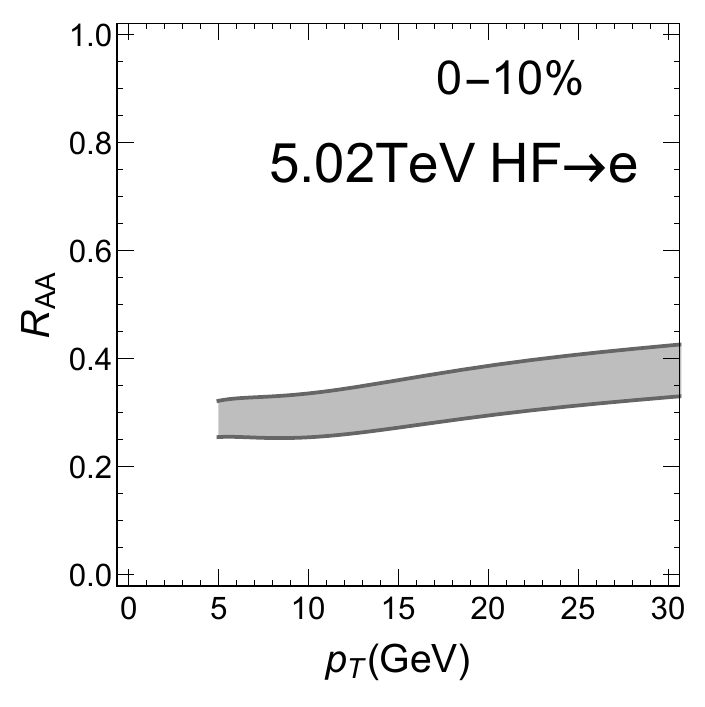}
\includegraphics[width=0.498\textwidth]{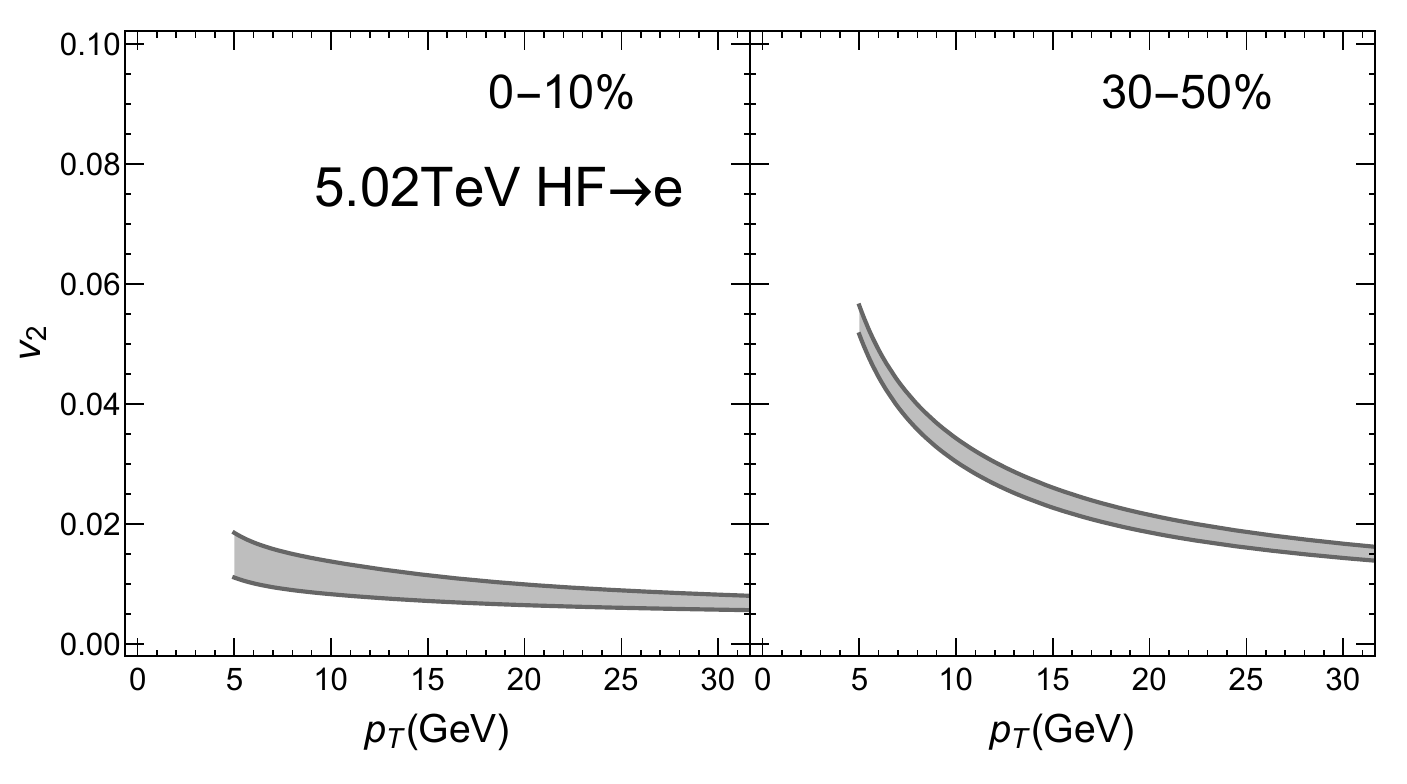}
\caption{$R_{AA}$(left) and $v_2$(right) for heavy flavor decayed muon in 5.02~TeV Pb-Pb collisions.}
\label{fig.pre_5020_e}
\end{figure}

Finally, we show the CUJET3 predictions for heavy flavor decayed muons and electrons in Figs.~\ref{fig.pre_5020_mu} and~\ref{fig.pre_5020_e}.
Being the decay product of both $D$ and $B$ mesons, the $R_{AA}$ in the lower $p_T$ regime is sensitive to relative ratios between $D$ and $B$ absolute cross sections.
We expect more stringent future tests from the heavy flavor sector to help further constrain CUJET3.

\clearpage
\section{Jet Transport Coefficient and Shear Viscosity}\label{sec.coef}

As discussed above, the jet quenching observables of light hadrons provide stringent constraints on values of the jet energy loss parameters. 
Furthermore, the comparison between three different schemes, (i) sQGMP-$\chi_T^L$, (ii) sQGMP-$\chi_T^u$, and (iii) wQGP, shows the need of chromo-magnetic-monopole degrees of freedom, robustly with respect to current theoretical uncertainties on the temperature dependence of the quark liberation rate.
It is of great interest to further compare how the jet and bulk transport properties differ in these schemes,
as this paves the way for clarifying the temperature dependence of jet quenching and shear viscous transport properties based on available high $p_T$ data in high-energy A+A collisions.

The jet transport coefficient $\hat{q}$ characterizes the averaged transverse momentum transfer squared per mean free path~\cite{Burke:2013yra}.
For a quark jet (in the fundamental representation F) with initial energy $E$, we calculate its $\hat{q}$ in the same way as the previous CUJET3.0 computation in~\cite{Xu:2014tda,Xu:2015bbz}, via
\begin{eqnarray}
\hat{q}_F(E,T)= & \int_0^{6ET} & dq_\perp^2  \frac{ 2\pi }{(\boldsymbol{q}_\perp^2 + f_E^2 \mu^2(\boldsymbol{z}))(\boldsymbol{q}_\perp^2 + f_M^2 \mu^2(\boldsymbol{z}))} \rho(T) \nonumber\\
&\times& \Big\{ \left[C_{qq} f_q + C_{qg} f_g \right]\cdot\left[  \alpha_s^2(\boldsymbol{q}_\perp^2)\right]\cdot\left [f_E^2 \boldsymbol{q}_\perp^2+ {f_E^2 f_M^2 \mu^2(\boldsymbol{z})} \right ] + \nonumber\\
&  &\;\;\; \left[ C_{qm} ( 1- f_q - f_g )\right]\cdot \left[ 1\right]\cdot \left[f_M^2{\boldsymbol{q}_\perp^2} + {f_E^2 f_M^2 \mu^2(\boldsymbol{z})}\right]  \Big\}\;,
\label{qhatF}
\end{eqnarray}
and similarly for a gluon/cmm jet:
\begin{eqnarray}
\hat{q}_{g}(E,T)= & \int_0^{6ET} & dq_\perp^2  \frac{ 2\pi }{(\boldsymbol{q}_\perp^2 + f_E^2 \mu^2(\boldsymbol{z}))(\boldsymbol{q}_\perp^2 + f_M^2 \mu^2(\boldsymbol{z}))} \rho(T) \nonumber\\
&\times& \Big\{ \left[C_{gq} f_q + C_{gg} f_g \right]\cdot\left[  \alpha_s^2(\boldsymbol{q}_\perp^2)\right]\cdot\left [f_E^2 \boldsymbol{q}_\perp^2+ {f_E^2 f_M^2 \mu^2(\boldsymbol{z})} \right ] + \nonumber\\
&  &\;\;\; \left[ C_{gm} ( 1- f_q - f_g )\right]\cdot \left[ 1\right]\cdot \left[f_M^2{\boldsymbol{q}_\perp^2} + {f_E^2 f_M^2 \mu^2(\boldsymbol{z})}\right]  \Big\}\;,
\label{qhatG}\\
\hat{q}_{m}(E,T)= & \int_0^{6ET} & dq_\perp^2  \frac{ 2\pi }{(\boldsymbol{q}_\perp^2 + f_E^2 \mu^2(\boldsymbol{z}))(\boldsymbol{q}_\perp^2 + f_M^2 \mu^2(\boldsymbol{z}))} \rho(T) \nonumber\\
&\times& \Big\{ \left[C_{mq} f_q + C_{mg} f_g \right]\cdot \left[ 1\right] \cdot\left [f_E^2 \boldsymbol{q}_\perp^2+ {f_E^2 f_M^2 \mu^2(\boldsymbol{z})} \right ] + \nonumber\\
&  &\;\;\; \left[ C_{mm} ( 1- f_q - f_g )\right]\cdot \left[ \alpha_s^{-2}(\boldsymbol{q}_\perp^2)\right] \cdot \left[f_M^2{\boldsymbol{q}_\perp^2} + {f_E^2 f_M^2 \mu^2(\boldsymbol{z})}\right]  \Big\}\;.
\label{qhatM}
\end{eqnarray} 
The quasi-parton density fractions of quark (q) or gluon (g), denoted as $f_{q,g}$, are defined as
\begin{eqnarray}\begin{split}
f_q&= c_q L(T) ,\;f_g=c_g L(T)^2,\qquad \;({\rm if}\;\chi_T^L) \\ 
f_q&= c_q \tilde{\chi}_2^u(T) ,\;f_g=c_g L(T)^2,\qquad ({\rm if}\;\chi_T^u) 
\end{split}\label{FracScheme}
\end{eqnarray}
respectively for sQGMP $\chi_T^L$ and $\chi_T^u$ scheme.
The magnetically charged quasi-particle density fraction is hence $f_m=1-\chi_T=1-f_q-f_g$. 
The color factors are given by 
\begin{eqnarray}\begin{split}
C_{qq} &= \frac{4}{9} , \; C_{gg}=C_{mm}=C_{gm}=C_{mg}=\frac{9}{4}  , \;\\
C_{qg} &= C_{gq} = C_{qm} = C_{mq} = 1 \;.
\end{split}\label{qhat1}
\end{eqnarray}
While switching to the wQGP scheme, by taking $f_q=c_q,\;f_g=c_g,\;f_E=1,\;f_M=0$, turning off the cmm channel, and employing the running coupling $\alpha_s(Q^2)$ defined in Eq.(\ref{AlphaRunMax}), the jet transport coefficient $\hat q$ for a quark/gluon jet defined in Eq.(\ref{qhatF}/\ref{qhatG}) returns to that of the CUJET2.0 framework~\cite{Xu:2014ica}.

\begin{figure}[!hbt]\centering
\includegraphics[width=0.4\textwidth]{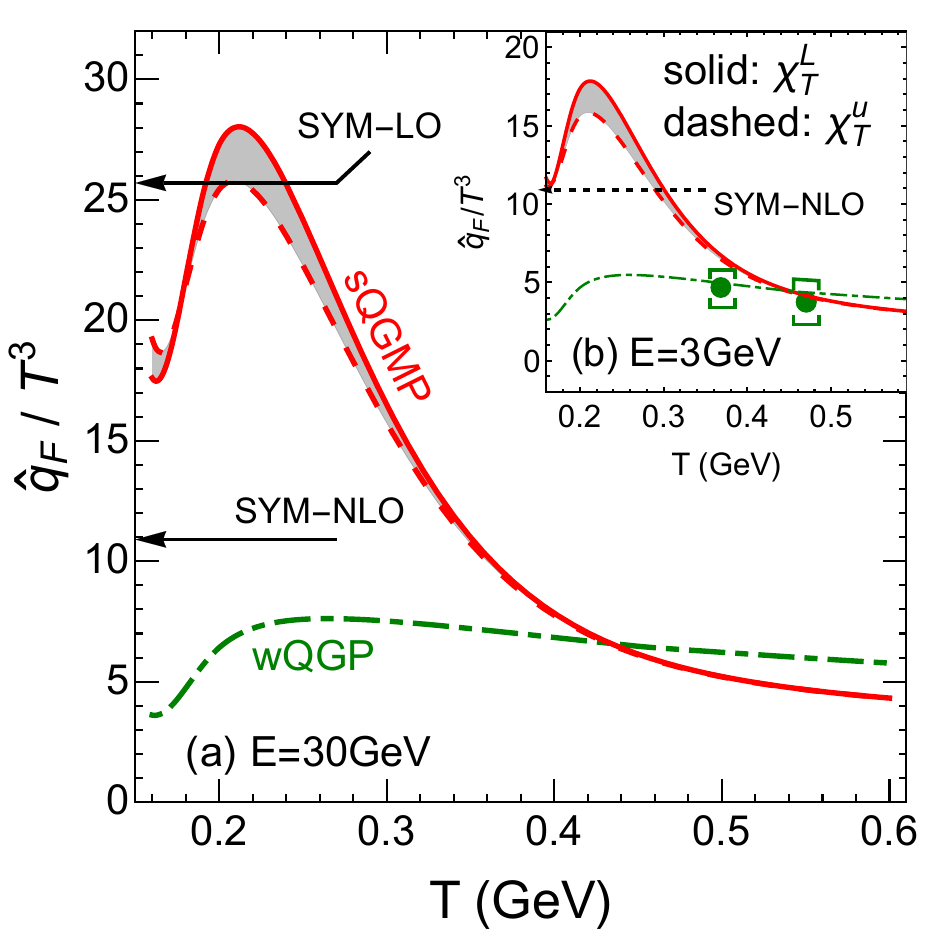}
\includegraphics[width=0.4\textwidth]{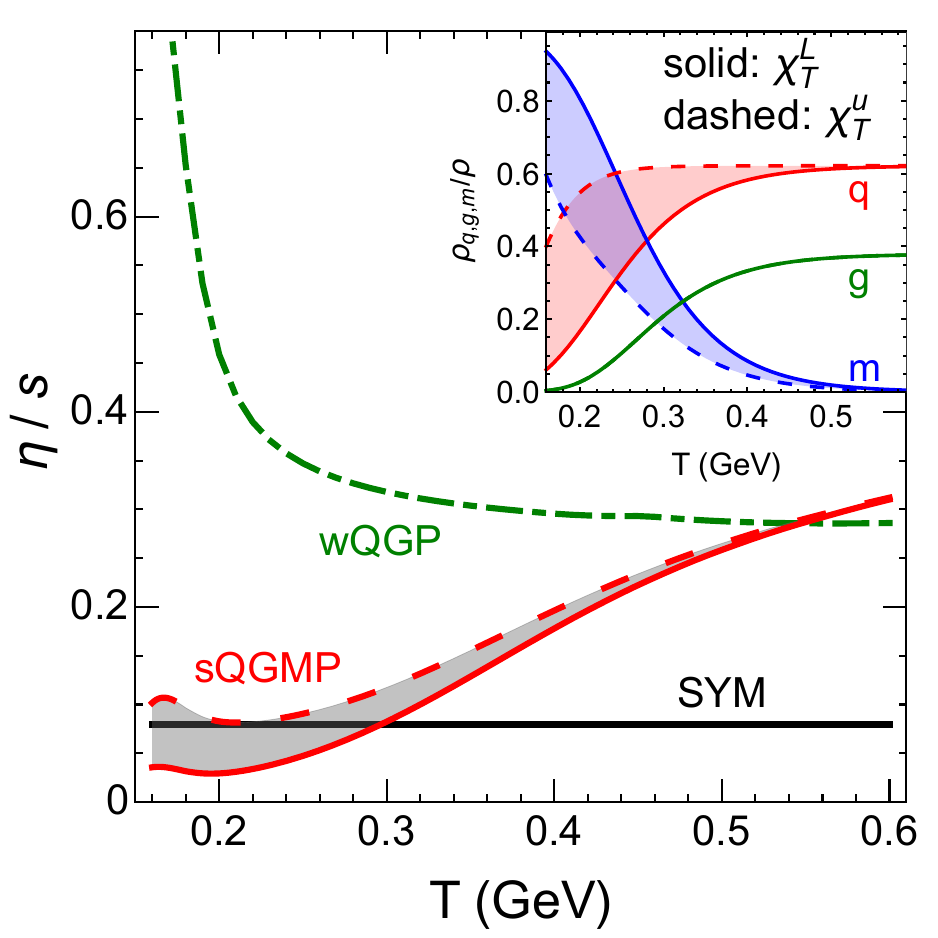}
\caption{\label{fig:FL-qhat-etas} 
(Color online)
(Left) Temperature dependence of the dimensionless jet transport coefficient $\hat{q}_F/T^3$ for a light quark jet with initial energy $E=$ (a) 30GeV, (b) 3GeV in CUJET framework with three schemes: (i) sQGMP-$\chi_T^L$ scheme (red solid curve), (ii) sQGMP-$\chi_T^u$ scheme (red dashed curve), and (iii) wQGP/CUJET2.0 scheme (green dotted-dashed curve). 
$\mathcal{N}=4$ leading order/next to leading order Super Yang-Mills 
$\hat{q}_{SYM-LO}/T^3=\frac{\pi^{3/2}\Gamma(\frac{3}{4})}{\Gamma(\frac{5}{4})}\sqrt{\lambda}$ and 
$\hat{q}_{SYM-NLO}/T^3=\frac{\pi^{3/2}\Gamma(\frac{3}{4})}{\Gamma(\frac{5}{4})}\sqrt{\lambda}(1-\frac{1.957}{\sqrt{\lambda}})$ respectively~\cite{Liu:2006ug} with coupling $\lambda=4\pi\cdot3\cdot0.31$, are plotted for comparison.
Green blobs in inset (b) shows the JET collaboration~\cite{Burke:2013yra} model average of $\hat{q}_F/T^3$ while boxes represent uncertainties.
(Right) Shear viscosity to entropy density ratio $\eta/s$ estimated with scheme (i) (red solid curve), (ii) (red dashed curve), and (iii) (green dotted-dashed curve). The inset shows quasi-particle number density fraction of q, g, m in liberation scheme $\chi_T^L$ (solid curve) and $\chi_T^u$ (dashed curve).
}
\end{figure}

Once the jet transport coefficient $\hat{q}$ has been computed, $\hat{q}(T,E)$ can be extrapolated down to thermal energy scales $E\sim 3T/2$ and the shear viscosity can be estimated to entropy density ratio $\eta/s$, based on kinetic theory in a weakly-coupled quasi-particle scenario~\cite{Danielewicz:1984ww,Hirano:2005wx,Majumder:2007zh}. 
An estimate of $\eta/s$ can be derived as
\begin{eqnarray}
\eta/s &=& \frac{1}{s}\, \frac{4}{15} 
\sum_{a} \rho_a \langle p\rangle_a \lambda_a^{\perp} \nonumber\\
&=&\frac{4T}{5s}  \sum_a \rho_a \left(\sum_b \rho_b \int_0^{\langle \mathcal{S}_{ab}  \rangle /2}dq_\perp^2 
\frac{4q_\perp^2}{\langle \mathcal{S}_{ab} \rangle }\frac{d\sigma_{ab}}{dq_\perp^2}\right)^{-1} \nonumber\\
&=&\frac{18T^3}{5s}  \sum_a \rho_a/{\hat{q}}_a(T,E=3T/2)\;\;.
\label{Effetas1} 
\end{eqnarray}
The $\rho_a(T)\equiv f_a\;\rho(T)$ is the quasi-parton density of type $a=q,g,m$. The mean thermal Mandelstam variable $\langle \mathcal{S}_{ab} \rangle \sim 18T^2$. Clearly the $\eta/s$ of the system is dominated by the ingredient which has the largest $\rho_a/\hat{q}_a$.

In the left panel of Fig.~\ref{fig:FL-qhat-etas}, we show the temperature dependence of the dimensionless jet transport coefficient $\hat{q}_F/T^3$ for a light quark jet with initial energy $E=$ 30GeV / 3GeV with all three schemes. Corresponding results from JET collaboration~\cite{Burke:2013yra} model average and AdS/CFT limit~\cite{Liu:2006ug} are also plotted for comparisons. As discussed in previous CUJET3.0 papers~\cite{Xu:2014tda,Xu:2015bbz}, the near-$T_c$ enhancement of dimensionless jet transport coefficient can be observed with robust dependence on quark liberation schemes.

In the right panel of Fig.~\ref{fig:FL-qhat-etas}, we show the shear viscosity to entropy density ratio $\eta/s$ estimated by kinetic theory using the $\hat{q}$ extrapolation Eq.~\eqref{Effetas1} with schemes (i) (red solid curve), (ii) (red dashed curve), and (iii) (green dotted-dashed curve). 
The inset shows the quasi-particle number density fraction of q, g, and m in the liberation scheme $\chi_T^L$ (solid curve) and $\chi_T^u$ (dashed curve). In the near $T_c$ regime, in the $\chi_T^u$ scheme,  the total $\eta/s$ is dominated by q, while in the $\chi_T^L$ ``slow'' quark liberation scheme the total $\eta/s$ is dominated by m. 
For each sQGMP scheme, there is a clear $\eta/s$ minimum at $T\sim210$~MeV, which is comparable with the SYM limit $(\eta/s)_{min}=1/4\pi$.

\section{Summary}\label{sec.summary}

In this study, we presented the CUJET3.1 framework and  performed a global quantitative $\chi^2$ analysis by comparsion with  a large set of light hadron jet quenching observables for central and semi-central heavy-ion collisions for beam energies $\sqrt{s_{NN}}=$ 200~GeV(Au-Au), 2.76~TeV(Pb-Pb), and 5.02~TeV(Pb-Pb). This analysis allows the optimization of the two key parameters in the CUJET3.1 framework, and the global $\chi^2$ is found to be minimized to near unity for
    $\alpha_c \approx 0.9\pm 0.1$, and $c_m\approx 0.25 \pm 0.03$. 
With such parameters,  the CUJET3 framework gives a unified, systematic and successful description of a comprehensive set of available data, from average suppression to azimuthal anisotropy, from light to heavy flavors, and from central to semi-peripheral collisions for all three colliding systems.
Thus, CUJET3.1 provides a non-perturbative solution to the long standing hard ($R_{AA}$ and $v_2$) versus soft ``perfect fluidity'' puzzle. 
Such a quantitative analysis strongly supports the necessity of including the interaction between jet and chromo-magnetic-monopoles to provide a consistent description of both $R_{AA}$ and $v_2$ across centrality and beam energy.

In this work, we also present CUJET3 predictions for a number of observables for additional tests.
We expect that the comparison between the light hadron $R_{AA}$ in 5.44~TeV Xe-Xe collisions and those observed in 5.02~TeV Pb-Pb collisions, will further test the path length dependence of the CUJET3 jet energy loss model.
The mass dependence of the jet energy loss in CUJET3 can also be further  tested   by its predictions  for $B$-decayed $D$ meson $R_{AA}$ in 5.02~TeV Pb-Pb collisions, to be compared with future precise measurement of this observable. 

Finally, we emphasize the important theoretical advantage of  the CUJET3.1 framework.
It is not only $\chi^2$ consistent with soft and hard observables data at RHIC and LHC, but also with non-perturbative lattice QCD data.
Remarkably, estimates from this framework lead to a shear viscosity to entropy density ratio $\frac{\eta}{s} \sim 0.1$, which are not only consistent with the extracted values from experimental soft + hard A + A phenomenology, but also theoretically internally consistent with the sQGMP kinetic theory link, $\frac{\eta}{s} \sim \frac{T^3}{\hat{q}_F(E \rightarrow 3 T, T)}$, between long distance collective fluid properties and short distance jet quenching physics especially near $T_c$.

\vspace{1cm}
\acknowledgments{
The authors are particularly grateful to Jiechen Xu for major contributions in establishing the CUJET3 framework. SS and JL are partly supported  by the National Science Foundation under Grant No. PHY-1352368. 
MG acknowledges support from IOPP of CCNU, Wuhan China. 
The computations in this study were performed on IU's Big Red II clusters, that are supported in part by Lilly Endowment, Inc., through its support for the Indiana University Pervasive Technology Institute, and in part by the Indiana METACyt Initiative. The Indiana METACyt Initiative at IU was also supported in part by Lilly Endowment, Inc.
} 

\vspace{1cm}

\begin{appendix}

\section{CUJET3 Framework}\label{sec.cujet3}
The CUJET3 model is a jet energy loss simulation framework built on a non-perturbative microscopic model for hot medium as semi-quark-gluon-monopole plasma (sQGMP), which integrates two essential elements of confinement, i.e. the suppression of quarks/gluons and emergent magnetic monopoles.
A detailed description of its framework can be found in previously published CUJET studies~\cite{Buzzatti:2011vt,Xu:2014ica,Burke:2013yra,Xu:2014tda,Xu:2015bbz}.
The CUJET3 model employs the TG elastic energy loss formula~\cite{Thoma:1990fm,Bjorken:1982tu,Peigne:2008nd} for collisional processes, with the energy loss given by 
\begin{equation}
\label{rcCUJETElastic}
\begin{split}
\frac{dE(\boldsymbol{z})}{d\tau}= & - C_R \pi \left[ \alpha_s(\mu(\boldsymbol{z}))\alpha_s( 6 E(\boldsymbol{z})\Gamma(\boldsymbol{z}) T(\boldsymbol{z})) \right] T(\boldsymbol{z})^2 \Big( 1+\frac{N_f}{6} \Big) \\
& \times \log \left[ \frac{6T(\boldsymbol{z})\sqrt{E(\boldsymbol{z})^2\Gamma(\boldsymbol{z})^2-M^2}}{\Big( E(\boldsymbol{z})\Gamma(\boldsymbol{z})-\sqrt{E(\boldsymbol{z})^2\Gamma(\boldsymbol{z})^2-M^2}+6T(\boldsymbol{z})\Big)\mu(\boldsymbol{z})} \right]\;,
\end{split}
\end{equation}
and the average number of collisions
\begin{equation}
\bar{N_c} = \int_{0}^{\tau_{max}} d\tau \left[ \frac{\alpha(\mu(\boldsymbol{z})) \alpha(6 E(\boldsymbol{z})\Gamma(\boldsymbol{z}) T(\boldsymbol{z}))}{\mu(\boldsymbol{z})^2} \right] \left[ \frac{\Gamma(\boldsymbol{z})}{\gamma_f} \frac{18 \zeta(3)}{\pi} (4+N_f) T(\boldsymbol{z})^3 \right]\;,
\label{rcNumOfColl}
\end{equation}
where the $E(\boldsymbol{z})$ integral equation is solved recursively.
For radiational processes, the CUJET3 model employs the dynamical DGLV opacity expansion theory~\cite{Gyulassy:1993hr,Gyulassy:2000er,Djordjevic:2003zk,Djordjevic:2008iz} with the Liao--Shuryak chromo-magnetic-monopole scenario~\cite{Liao:2006ry,Liao:2007mj,Liao:2008jg,Liao:2008dk,Liao:2012tw}. The inclusive single gluon emission spectrum at $n=1$ opacity series reads:
\begin{eqnarray}
\begin{split}
 x_E\frac{dN_g^{n=1}}{dx_E} =&\; \frac{18 C_R}{\pi^2} \frac{4+N_f}{16+9N_f} \int{d\tau}\; \rho(\boldsymbol{z}) \Gamma(\boldsymbol{z})\;\int{d^2k_{\perp}} \alpha_s\Big(\frac{\boldsymbol{k}_{\perp}^2}{x_+ (1-x_+)}\Big)\;\\
&\times\;\int{d^2q}\frac{ \alpha_s^2(\boldsymbol{q}_\perp^2)\left (f_E^2+ \frac{f_E^2 f_M^2 \mu^2(\boldsymbol{z})}{\boldsymbol{q}_\perp^2} \right )\chi_T+\left (f_M^2+ \frac{f_E^2 f_M^2 \mu^2(\boldsymbol{z})}{\boldsymbol{q}_\perp^2} \right) (1- \chi_T) }{(\boldsymbol{q}_\perp^2 + f_E^2 \mu^2(\boldsymbol{z}))(\boldsymbol{q}_\perp^2 + f_M^2 \mu^2(\boldsymbol{z}))} \\
&\times\;\frac{-2(\boldsymbol{k}_{\perp}-\boldsymbol{q}_{\perp})}{(\boldsymbol{k}_{\perp}-\boldsymbol{q}_{\perp})^2+\chi^2(\boldsymbol{z})} \left[ \frac{\boldsymbol{k}_{\perp}}{\boldsymbol{k}_{\perp}^2+\chi^2(\boldsymbol{z})} - \frac{(\boldsymbol{k}_{\perp}-\boldsymbol{q}_{\perp})}{(\boldsymbol{k}_{\perp}-\boldsymbol{q}_{\perp})^2+\chi^2(\boldsymbol{z})} \right] \\
&\times\;{\left[1-\cos\Big(\frac{(\boldsymbol{k}_{\perp}-\boldsymbol{q}_{\perp})^2+\chi^2(\boldsymbol{z})}{2 x_+ E } \tau\Big)\right]}\left(\frac{x_E}{x_+}\right)\left| \frac{dx_+}{dx_E} \right| \;\;.
\end{split}
\label{emEnergyLoss}
\end{eqnarray}
$C_R=4/3$ or $3$ is the quadratic Casimir of the quark or gluon; 
the transverse coordinate of the hard parton is denoted by $ \boldsymbol{z}=\Big( x_0+\tau\cos\phi,y_0+\tau\sin\phi; \tau\Big)$;
$E$ is the energy of the hard parton in the lab frame;
$\bf{k}_{\perp}$ ($|{\bf k}_{\perp}|\leq x_EE\cdot\Gamma(\boldsymbol{z})$) and $\bf{q}_{\perp}$ ($|{\bf q}_{\perp}|\leq 6T(\boldsymbol{z})E\cdot\Gamma(\boldsymbol{z})$) are the local transverse momentum of the radiated gluon and the local transverse momentum transfer, respectively. 
The gluon fractional energy $x_E$ and fractional plus-momentum $x_+$ are connected by $x_+(x_E)=x_E[1+\sqrt{1-(k_\perp/x_EE)^2}]/2$. 
We note that in the temperature range $T\sim T_c$, the coupling $\alpha_s$ becomes non-perturbative~\cite{Liao:2006ry,Liao:2008jg,Zakharov:2008kt,Randall:1998ra}.  Analysis of lattice data~\cite{Liao:2008jg} suggests the following thermal running coupling form:
\begin{equation}
\alpha_s(Q^2)=\dfrac{\alpha_c}{1+\frac{9\alpha_c}{4\pi} \log(\frac{Q^2}{T_c^2})}\;,
\label{TcEnhancement}
\end{equation}
with $T_c=160$~MeV. Note that at large $Q^2$, Eq.~\eqref{TcEnhancement} converges to vacuum running $\alpha_s(Q^2)=\frac{4\pi}{9\log(Q^2/\Lambda^2)}$, while at $Q=T_c$, $\alpha_s(T^2_c)=\alpha_c$.

The particle number density $\rho(\boldsymbol{z})$ is determined by the medium temperature $T(\boldsymbol{z})$ via $\rho(T) = \xi_s s(T)$, where $\xi_s = 0.253$ for $N_c=3$, $N_f=2.5$ Stefan-Boltzmann gas, and $s(T)$ is the bulk entropy density.
 In the presence of hydrodynamical four-velocity fields $u_f^{\mu}(\boldsymbol{z}) $, boosting back to the lab frame, a relativistic correction $\Gamma(\boldsymbol{z})=u^{\mu}_fn_{\mu}$ should be taken into account~\cite{Liu:2006he,Baier:2006pt}, where the flow four-velocity $u^{\mu}_f=\gamma_f(1,\vec{\beta}_f)$ and null hard parton four-velocity $n^\mu=(1,\vec{\beta}_{j})$. 
The bulk evolution profiles $(T(\boldsymbol{z}),\rho(\boldsymbol{z}),u_f^{\mu}(\boldsymbol{z}) )$ are generated from the VISH2+1 code~\cite{Song:2008si,Shen:2010uy,Renk:2010qx} with event-averaged Monte-Carlo Glauber initial condition, $\tau_0=0.6$ fm/c, s95p-PCE Equation of State (EOS), $\eta/s=0.08$, and Cooper-Frye freeze-out temperature 120~MeV~\cite{Song:2010mg,Majumder:2011uk,Qiu:2011hf,Shen:2011eg,Shen:2012vn,Shen:2014vra}. Event-averaged smooth profiles are embedded, and the path integrations $\int d\tau$ for jets initially produced at transverse coordinates $({\bf x}_0,\phi)$ are cut-off at dynamical $T(\boldsymbol{z}({\bf x}_0,\phi,\tau))|_{\tau_{max}}\equiv T_{cut} =160$~MeV hyper-surfaces~\cite{Xu:2014ica}.

In the CUJET2 framework, assuming weakly-coupling QGP, the running coupling takes the form (with $\Lambda_{QCD}=200$~MeV)
\begin{equation}
\alpha_s(Q^2) = \begin{cases}
\alpha_{\mathrm{max}} & \mbox{if } Q \le Q_{min}\;, \\
\dfrac{4\pi}{9\log(Q^2/\Lambda_{QCD}^2)}  & \mbox{if } Q > Q_{min}\;.
\end{cases}
\label{AlphaRunMax}
\end{equation}

The Debye screening mass $\mu(\boldsymbol{z})$ is determined from solving the self-consistent equation 
\begin{eqnarray}
\mu(\boldsymbol{z}) = \sqrt{4\pi\alpha_s(\mu^2(\boldsymbol{z}))}T(\boldsymbol{z})\sqrt{1+N_f/6}
\label{DebyeMass}
\end{eqnarray}
as in~\cite{Peshier:2006ah}; $\chi^2(\boldsymbol{z})=M^2 x_+^2+m_g^2(\boldsymbol{z})(1-x_+)$ regulates the soft collinear divergences in the color antennae and controls the Landau-Pomeranchuk-Migdal (LPM) phase, the gluon plasmon mass $ m_g(\boldsymbol{z})=f_E\mu(\boldsymbol{z}) / \sqrt{2} $. 

Because the sQGMP contains both chromo-electrically charged quasi-particles (cec) and chromo magnetically charged quasi-particles (cmc), when jets propagate through the medium near $T_c$, the total quasi-particle number density $\rho$ is divided into EQPs with fraction $\chi_T = \rho_E / \rho$ and MQPs with fraction $1-\chi_T=\rho_M/\rho$. The parameter $f_E$ and $f_M$ is defined via $f_E\equiv\mu_E/\mu$ and $f_M\equiv\mu_M/\mu$, with $\mu_E$ and $\mu_M$ being the electric and magnetic screening mass, respectively, following
\begin{eqnarray}
f_E(T(\boldsymbol{z})))  = \sqrt{\chi_T(T(\boldsymbol{z}))}  \;,\quad  \quad f_M(T(\boldsymbol{z})) = c_m \, g(T(\boldsymbol{z}))\, ,
\label{f_EM}
\end{eqnarray}
with the local electric ``coupling'' $g(T(\boldsymbol{z}))=\sqrt{4\pi\alpha_s(\mu^2(T(\boldsymbol{z})))}$.

In current sQGMP modeling, the cec component fraction $\chi_T$ remains a theoretical uncertainty related to the question of how fast the color degrees of freedom get liberated.
To estimate $\chi_T$, one notices that: (1) when temperature is high, $\chi_T$ should reach unity, i.e. $\chi_T(T\gg T_c) \to 1$;
(2) in the vicinity of the regime $T\sim (1-3)T_c$, the renormalized expectation value of the Polyakov loop L (let us redefine $L\equiv \ell = \langle \mathrm{tr} \mathcal{P} \exp\lbrace ig\int_{0}^{1/T} d\tau A_0\rbrace  \rangle /N_c$) deviates significantly from unity, implying the suppression $\sim L$ for quarks and $\sim L^2$ for gluons in the semi-QGP model~\cite{Hidaka:2008dr,Hidaka:2009ma,Dumitru:2010mj,Lin:2013efa}.
Consequently, in the liberation scheme ($\chi_T^L$-scheme), we define the cec component fraction as
\begin{equation}
\chi_T(T)  \equiv \chi_T^L(T)  = c_q L(T)  + c_g L^2(T) 
\end{equation}
for the respective fraction of quarks and gluons, where we take the Stefan-Boltzmann (SB) fraction coefficients, $c_q = (10.5 N_f )/(10.5 N_f + 16)$ and $c_g = 16/(10.5 N_f + 16)$, and the temperature dependent Polyakov loop $L(T)$ parameterized as ($T$ in~GeV)
\begin{eqnarray}
L(T) = \left[\frac{1}{2}+\frac{1}{2}{\rm Tanh}[7.69(T-0.0726)]\right]^{10} 
\label{PolyakovLoop},
\end{eqnarray}
adequately fitting both the HotQCD~\cite{Bazavov:2009zn} and Wuppertal-Budapest~\cite{Borsanyi:2010bp} lattice results.

On the other hand, another useful measure of the non-perturbative suppression of the color electric DOF is provided by the quark number susceptibilities~\cite{McLerran:1987pz,Gottlieb:1988cq,Gavai:1989ce,Gottlieb:1987ac}.
The diagonal susceptibility is proposed as part of the order parameter for chiral symmetry breaking/restoration in~\cite{McLerran:1987pz}, 
and plays a similar role as properly renormalized $L$ for quark DOFs.
In this scheme, we parametrize the lattice diagonal susceptibility of $u$ quark number density, renormalizing the susceptibility by its value at $T\rightarrow \infty$, as  ($T$ in~GeV)
\begin{eqnarray}
\tilde{\chi}_2^u(T) \equiv \frac{\chi_2^u(T)}{0.91} = \left[ \frac{1}{2} \left\lbrace 1+{\rm Tanh}[15.65(T-0.0607)] \right\rbrace  \right]^{10}
\label{chi2u_},
\end{eqnarray} 
and define the cec component fraction in the deconfinement scheme ($\chi_T^u$-scheme) as:
\begin{equation}
\chi_T(T) \equiv \chi_T^u(T)  = c_q \tilde\chi_2^u(T)  + c_g L^2(T) .
\end{equation}
These two different schemes, for the rate of ``quark liberation'', with $\chi_T^L$ the ``slow'' and $\chi_T^u$ the ``fast'', provide useful estimates of theoretical systematic uncertainties associated with the quark component of the sQGMP model.

\begin{figure}[!hbt]\centering
\includegraphics[width=0.43\textwidth]{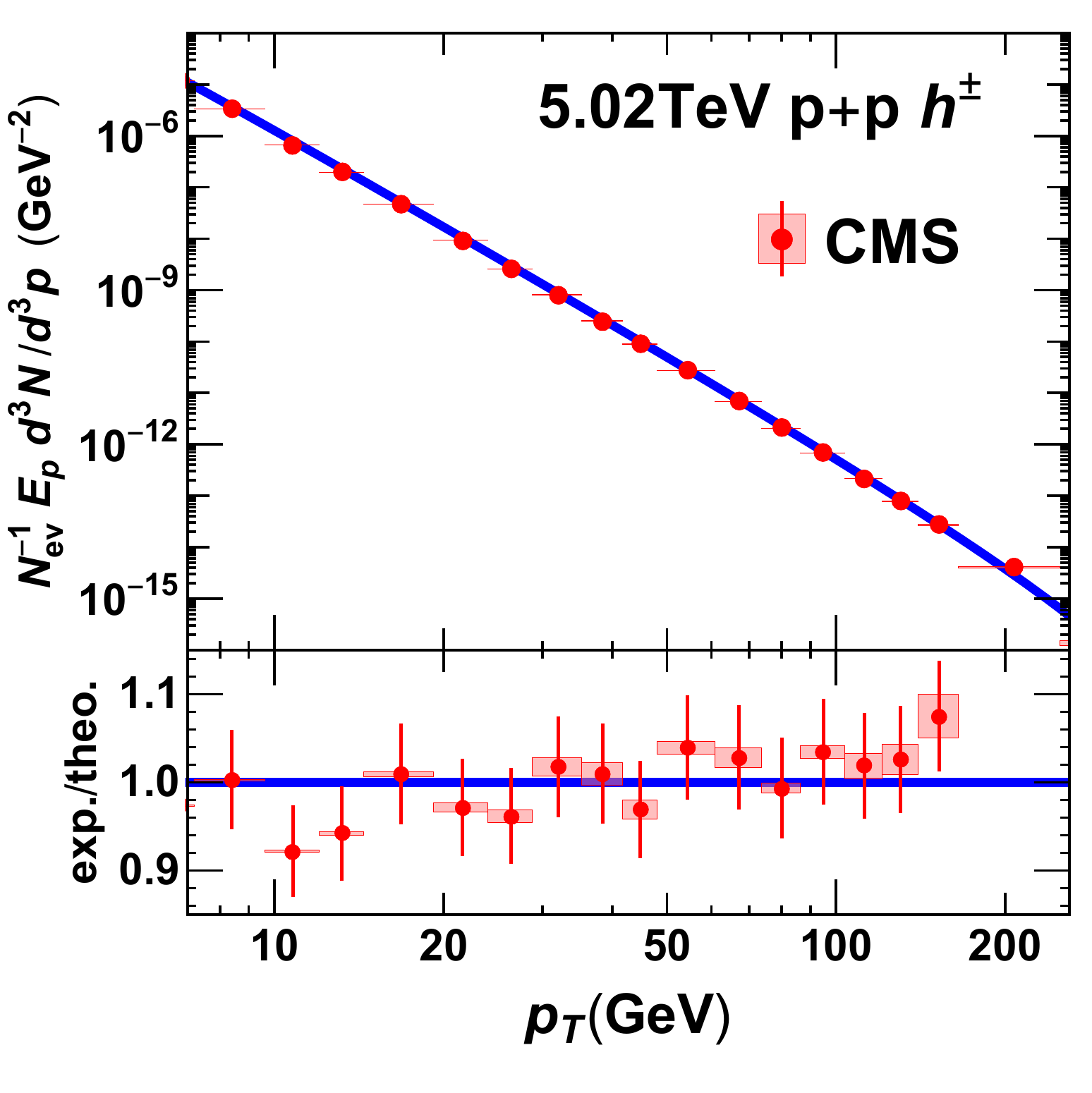}
\caption{(color online) Comparison between initial unquenched invariant momentum distribution of charged hadron, predicted using CTEQ5~{\protect\cite{Wang:private}} PDF and KKP fragmentation~{\protect\cite{Kniehl:2000fe}}, and 5.02 TeV CMS data within $|\eta|<1$~{\protect\cite{Khachatryan:2016odn}}. Experiment-to-theory ratio is also shown in the lower panel.
}
\label{fig.ppspectra}
\end{figure}
Finally, in the CUJET3 framework, the $p+p$ spectra of light quarks and gluons are generated by LO pQCD~\cite{Wang:private} calculations with CTEQ5 Parton Distribution Functions (PDF); while those of charm and bottom quarks are generated from the FONLL calculation~\cite{Cacciari:2005rk} with CTEQ6M Parton Distribution Functions.
In the meantime, the spectra of light hadrons are computed with KKP Fragmentation Functions~\cite{Kniehl:2000fe}; and those of open heavy flavor mesons are computed with Peterson Fragmentation Functions~\cite{Peterson:1982ak} (taking $\epsilon=0.06$ for D meson, and $\epsilon=0.006$ for B meson).
The decay of heavy flavor mesons into leptons, including $D \to \ell$, $B \to \ell$, and $B \to D \to \ell$ channels, follows the same parameterization as in~\cite{Cacciari:2005rk}. Comparison between theory predictions and experimental measurements on the initial unquenched invariant $p+p\rightarrow h^{\pm}+X$ distribution in shown in the right panel of Fig.~\ref{fig.ppspectra}, for $5.02$~TeV collisions, and in a previous CUJET3.0 study~\cite{Xu:2015bbz}.

\section{Improvements in CUJET3.1}
\begin{figure}[!hbt]\centering
\includegraphics[width=0.5\textwidth]{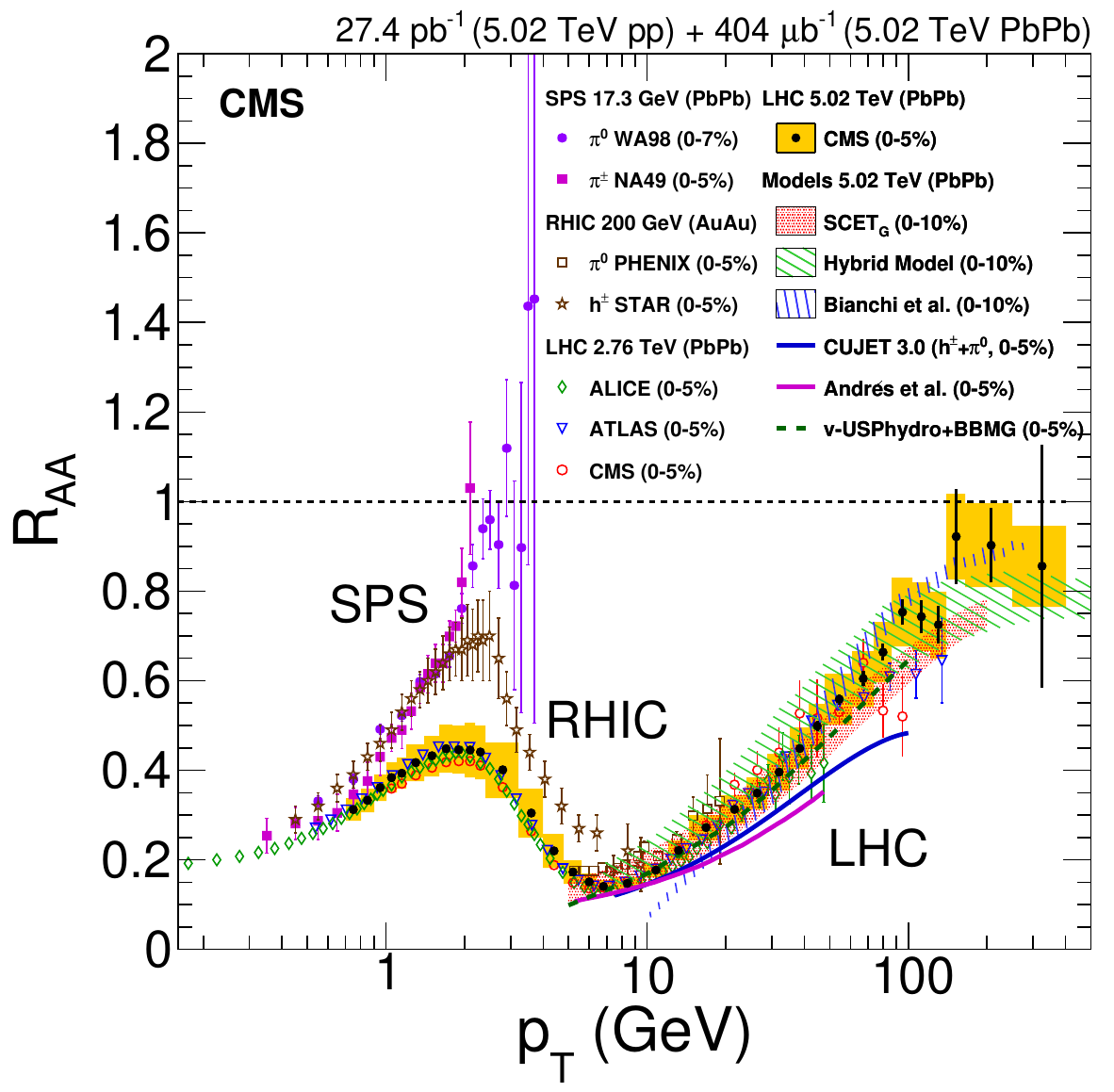}
\caption{(color online) Reproduced from Fig.~5 of CMS Ref.~{\protect~\cite{Khachatryan:2016odn}} (with permission): 
$R_{AA}$ results as a function of $p_T$ in {\protect (0 -- 5)\%} centrality class. 
Vertical bars (shaded boxes) represent statistical (systematic) uncertainties. 
Blue curve represents calculation made with CUJET3.0 {\protect~\cite{Xu:2015bbz}}s.}
\label{fig.CMS-Raa}
\end{figure}

\begin{figure}[!hbt]\centering
\includegraphics[width=0.8\textwidth]{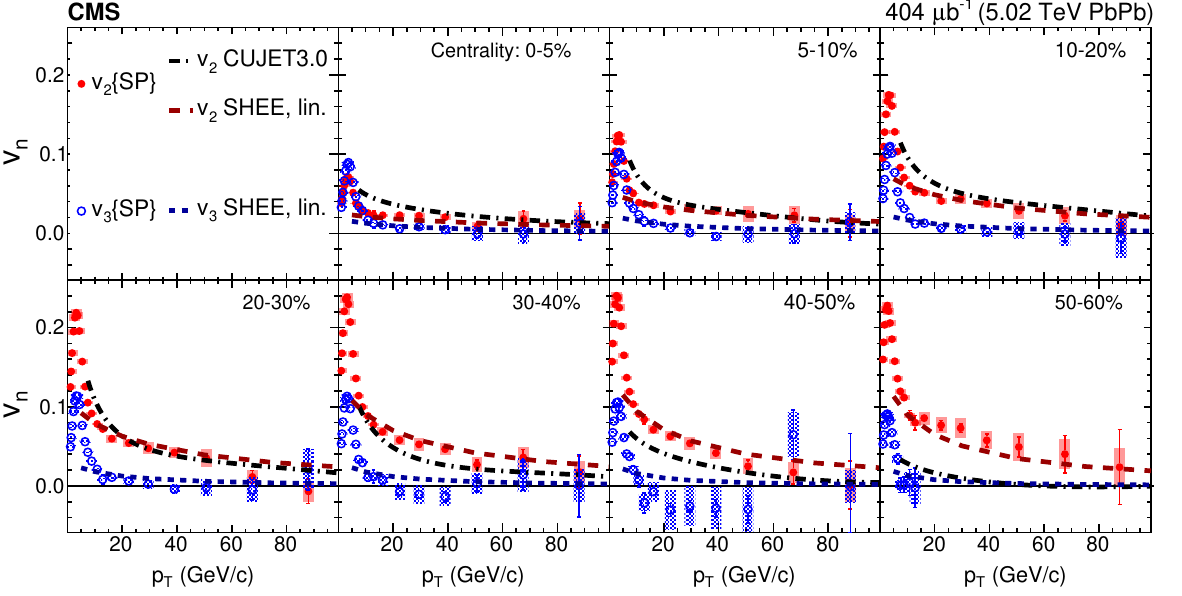}
\caption{(color online) Reproduced from Fig.~1 of CMS Ref.~{\protect~\cite{Sirunyan:2017pan}} (with permission): 
$v_2$ and $v_3$ results from SP method as a function of pT, in seven collision centrality ranges from {\protect (0 -- 5)\% to (50 -- 60)\%}. 
Vertical bars (shaded boxes) represent statistical (systematic) uncertainties.
Curves represent calculations made with CUJET 3.0~\cite{Xu:2015bbz} and SHEE~\cite{Noronha-Hostler:2016eow} models. }   
\label{fig.CMS-v2}
\end{figure}

In this Appendix, we discuss the improvements of the CUJET3.1 framework with respect to the earlier CUJET3.0 framework version. One important motivation for the CUJET3.1 upgrade reported in the present paper was to uncover causes and correct the discrepancy of CUJET3.0 predictions for LHC $5.02$~ATeV Pb + Pb collusions, reported by CMS in 
Ref.~\cite{Khachatryan:2016odn} with the nuclear modification factor $R_{AA}$ (see Fig.~\ref{fig.CMS-Raa}), as well as in Ref.~\cite{Sirunyan:2017pan} with their observed $p_T$ and especially the centrality dependence of the hard elliptic asymmetry $v_2$ (see Fig.~\ref{fig.CMS-v2}). 


After a systematic examination, we found and corrected two issues in previous CUJET3.0 simulations for 5.02~ATeV Pb + Pb collisions.
(i) First, the initial parton spectra were not consistently read in: the flavor factor of 3 was missed for light quark spectra. 
Resultabtly, a higher fraction of the final hadrons were fragmented from gluon jets, which are more quenched relative to quark jets and caused the over-quenched $R_{AA}$.
(ii) Second, the probability distribution of initial jet production was incorrectly oriented (with $x$- and $y$-axis switched), and consequently wrong centrality dependence of $v_2$ was predicted.
By correcting these two issues, the CUJET3.1 simulation correctly reproduces the $p_T$ and centrality dependence of both $R_{AA}$ and $v_2$.
Details of the comparison are shown in Figs.~\ref{fig.light_5020_RAA} and~\ref{fig.light_5020_v2} in Sec.~\ref{sec.comparison.light}.
\end{appendix}

\vspace{1cm}
\bibliography{Ref}
\end{CJK*}
\end{document}